\documentclass{ieeetj}

\def\BibTeX{{\rm B\kern-.05em{\sc i\kern-.025em b}\kern-.08em
    T\kern-.1667em\lower.7ex\hbox{E}\kern-.125emX}}
\AtBeginDocument{\definecolor{tmlcncolor}{cmyk}{0.93,0.59,0.15,0.02}\definecolor{NavyBlue}{RGB}{0,86,125}}

\def\OJlogo{\vspace{-4pt}$<$Society logo(s) and publication title will appear here.$>$}
\def\seclogo{\vspace{10pt}$<$Society logo(s) and publication title will appear here.$>$}

\def\authorrefmark#1{\ensuremath{^{\textbf{#1}}}}

\usepackage{graphicx, color}
\usepackage{xcolor}
\usepackage{amsmath,amssymb,amsfonts}
\usepackage{amsthm}
\usepackage{algorithmic}
\usepackage{algorithm}
\usepackage{array}
\usepackage[caption=false,font=normalsize,labelfont=sf,textfont=sf]{subfig}
\usepackage{mathtools}
\usepackage{textcomp}
\usepackage{stfloats}
\usepackage{url}
\usepackage{verbatim}
\usepackage{cite}
\usepackage{multirow}
\usepackage{colortbl}
\usepackage{hyperref}
\usepackage{booktabs}
\hypersetup{hidelinks=true}
\DeclareMathOperator{\prox}{prox}

\newcommand{\argmin}{\mathop{\mathrm{argmin}}\limits}

\def\InnerProduct<#1>{\langle #1 \rangle}
\def\MatrixBrackets[#1]#2{\lbrack #1 \rbrack_{#2}}
\def\AlgoLfloor#1{\left\lfloor #1 \right}

\def\OpNorm#1{\| #1 \|_{\mathrm{op}}}
\def\OpNormSq#1{\| #1 \|_{\mathrm{op}}^{2}}

\def\llHTV{{l_{0}\text{-}l_{1}\text{HTV}}}

\def\RealSpace#1{\mathbb{R}^{#1}}
\def\SetProperLowerConvex#1{\Gamma_{0}(#1)}
\def\SetConvex{C}
\def\FuncIndicator#1{\iota_{#1}}

\def\Projection#1{P_{#1}}

\def\FuncOne{f}
\def\FuncTwo{g}
\def\NumVarOne{N}
\def\NumVarTwo{M}
\def\VarOne{\mathbf{x}}
\def\VarTwo{\mathbf{y}}

\def\ElementOne#1{x_{#1}}
\def\IndexOne{i}

\def\HSIClean{\mathbf{u}}

\def\HSIObsv{\mathbf{v}}
\def\NoiseSparse{\mathbf{s}}
\def\NoiseGauss{\mathbf{n}}
\def\NoiseStripe{\mathbf{t}}

\def\ResNoiseSparse{\NoiseSparse^{'}}
\def\ResNoiseStripe{\NoiseStripe^{'}}

\def\NumElement{N}
\def\NumVert{N_{1}}
\def\NumHori{N_{2}}
\def\NumBand{N_{3}}

\def\IndexVert{n_{1}}
\def\IndexHori{n_{2}}

\def\IndexAlg{t}

\def\NotaVector{\mathbf{x}}
\def\NotaMatrix{\mathbf{X}}
\def\NotaScalar{x}
\def\NotaNumDim{N}
\def\NotaIndex{n}

\def\FuncProx{\FuncOne}
\def\CoefProx{\gamma}
\def\VarProxOne{\VarOne}
\def\VarProxTwo{\VarTwo}
\def\NumVarProx{\NumVarOne}

\def\Projection#1{P_{#1}}

\def\FuncPrimal#1{\FuncOne_{#1}}
\def\FuncDual#1{\FuncTwo_{#1}}
\def\VarPrimal#1{\VarOne_{#1}}
\def\VarDual#1{\VarTwo_{#1}}

\def\NumVarPrimal{\NumVarOne}
\def\NumVarDual{\NumVarTwo}
\def\DimVarPrimal#1{n_{#1}}
\def\DimVarDual#1{m_{#1}}
\def\IndexPrimal{i}
\def\IndexDual{j}

\def\LinOpPPDS#1{\mathbf{A}_{#1}}
\def\ScalarStepsize#1{\gamma_{#1}}

\def\ParamStepsize#1{\gamma_{#1}}

\def\OpIpoVert{\mathbf{L}_{\updownarrow}}
\def\OpIpoHori{\mathbf{L}_{\leftrightarrow}}
\def\OpIpoCent{\mathbf{L}_{\bullet}}
\def\OpIpo{\mathbf{L}}
\def\OpIpoVertT{\OpIpoVert^{\top}}
\def\OpIpoHoriT{\OpIpoHori^{\top}}
\def\OpIpoCentT{\OpIpoCent^{\top}}
\def\OpIpoT{\OpIpo^{\top}}

\def\OpIpoCTV{\OpIpo_{g}}
\def\OpIpoCTVT{\OpIpo_{g}^{\top}}
\def\DiffOpSpCTV{\mathbf{D}_{g}}

\def\VarCTVImage{\mathbf{x}}
\def\VarCTVDual{\mathbf{y}}
\def\VarCTVDualVert{\VarCTVDual_{v}}
\def\VarCTVDualHori{\VarCTVDual_{h}}

\def\VarCTVGrad{\mathbf{w}}
\def\VarCTVGradVert{\VarCTVGrad_{\updownarrow}}
\def\VarCTVGradHori{\VarCTVGrad_{\leftrightarrow}}
\def\VarCTVGradCent{\VarCTVGrad_{\bullet}}

\def\VarCTVGradVertT{\VarCTVGradVert^{\top}}
\def\VarCTVGradHoriT{\VarCTVGradHori^{\top}}
\def\VarCTVGradCentT{\VarCTVGradCent^{\top}}

\def\CondatTV{\text{TV}_{\text{c}}}

\def\GeoSSTV{\text{GeoSSTV}}
\def\IpoOne{\mathbf{w}_{1}}
\def\IpoTwo{\mathbf{w}_{2}}
\def\ParamBalance{\omega}

\def\IndexBand{\IndexPrimal}

\def\DiffOpVert{\mathbf{D}_{v}}
\def\DiffOpHori{\mathbf{D}_{h}}
\def\DiffOpBand{\mathbf{D}_{s}}
\def\DiffOpSp{\mathbf{D}}

\def\DiffOpVertT{\mathbf{D}_{v}^{\top}}
\def\DiffOpHoriT{\mathbf{D}_{h}^{\top}}
\def\DiffOpBandT{\mathbf{D}_{s}^{\top}}
\def\DiffOpSpT{\mathbf{D}^{\top}}

\def\RadiusFidel{\varepsilon}
\def\RadiusSparse{\alpha}
\def\RadiusStripe{\beta}

\def\ParamsRadius{\rho}

\def\BallFidel{B_{2, \RadiusFidel}^{\HSIObsv}}
\def\BallSparse{B_{1, \RadiusSparse}}
\def\BallStripe{B_{1, \RadiusStripe}}
\def\SetZero{\{ \mathbf{0} \}}

\def\MinRange{\underline{\mu}}
\def\MaxRange{\bar{\mu}}
\def\SetRange{R_{\MinRange,\MaxRange}}

\def\GroupIndex{g}

\def\IndexProx{i}

\def\BallFidel{B_{2, \RadiusFidel}^{\HSIObsv}}
\def\BallSparse{B_{1, \RadiusSparse}}
\def\BallStripe{B_{1, \RadiusStripe}}
\def\SetZero{\{ \mathbf{0} \}}
\def\MinRange{\underline{\mu}}
\def\MaxRange{\bar{\mu}}
\def\SetRange{R_{\MinRange,\MaxRange}}

\def\IndexVert{i}
\def\IndexHori{j}

\def\HSICleanRes{\HSIClean^{'}}

\def\IpoOneRes{\IpoOne^{'}}
\def\IpoTwoRes{\IpoTwo^{'}}

\def\StanDevGauss{\sigma}
\def\RateSparse{p_{\NoiseSparse}}
\def\RateStripe{p_{\NoiseStripe}}
\def\RateDeadline{p_{\text{d}}}
\def\CoverRateDeadline{c_{\text{d}}}

\def\MeanHSIObsv{\mu_{\HSIObsv}}

\begin{document}
\bstctlcite{IEEEexample:BSTcontrol}
\receiveddate{XX Month, XXXX}
\reviseddate{XX Month, XXXX}
\accepteddate{XX Month, XXXX}
\publisheddate{XX Month, XXXX}
\currentdate{XX Month, XXXX}
\doiinfo{XXXX.2022.1234567}

\markboth{}{Author {et al.}}

\title{Geometric Spatio-Spectral Total Variation for Hyperspectral Image Denoising and Destriping}

\author{Shingo Takemoto\authorrefmark{1}, Student Member, IEEE, and Shunsuke Ono\authorrefmark{2}, Senior Member, IEEE}
\affil{Department of Computer Science, Institute of Science Tokyo, Yokohama 226-8501, Japan}
\affil{Department of Computer Science, Institute of Science Tokyo, Yokohama 226-8501, Japan}
\corresp{Corresponding author: Shingo Takemoto (email: takemoto.s.e908@m.isct.ac.jp).}
\corresp{Corresponding author: Shunsuke Ono (email: ono.s.5af2@m.isct.ac.jp).}
\authornote{This work was supported in part by Grant-in-Aid for JSPS Fellows Grant Number JP24KJ1068, in part by JSPS KAKENHI under Grant 22H03610, 22H00512, 23H01415, 23K17461, 24K03119, 24K22291, 25H01296, and 25K03136, and in part by JST FOREST under Grant JPMJFR232M and JST AdCORP under Grant JPMJKB2307.}

\begin{abstract}
This article proposes a novel regularization method, named Geometric Spatio-Spectral Total Variation (GeoSSTV), for hyperspectral (HS) image denoising and destriping. HS images are inevitably affected by various types of noise due to the measurement equipment and environment. Total Variation (TV)-based regularization methods that model the spatio-spectral piecewise smoothness inherent in HS images are promising approaches for HS image denoising and destriping. However, existing TV-based methods are based on classical anisotropic and isotropic TVs, which cause staircase artifacts and lack rotation invariance, respectively, making it difficult to accurately recover round structures and oblique edges. To address this issue, GeoSSTV introduces a geometrically consistent formulation of TV that measures variations across all directions in a Euclidean manner. Through this formulation, GeoSSTV removes noise while preserving round structures and oblique edges. Furthermore, we formulate the HS image denoising problem as a constrained convex optimization problem involving GeoSSTV and develop an efficient algorithm based on a preconditioned primal-dual splitting method. Experimental results on HS images contaminated with mixed noise demonstrate the superiority of the proposed method over existing approaches.

\end{abstract}

\begin{IEEEkeywords}
denoising, destriping, hyperspectral image, rotation invariance, total variation
\end{IEEEkeywords}


\maketitle

\section{Introduction}
\label{sec:Introduction}
\IEEEPARstart{H}{yperspectral} (HS) imaging acquires hundreds of nearly contiguous spectral bands for each pixel, providing rich spectral signatures that can distinguish materials and enable quantitative analysis, which conventional RGB images cannot achieve. Owing to this unique capability, HS images have been widely applied in diverse fields such as remote sensing, environmental monitoring, agriculture, and mineral exploration~\cite{Borengasser2007HSIApplications,Grahn2007Techniques, Thenkabail2016VegetationOverview,Lu2020AgricultureOverview}. However, HS data are inevitably contaminated by noise during the imaging process arising from physical and instrumental factors, such as sensor thermal effects, photon shot noise, and defective or dead pixels~\cite{Rasti2018DenoisingOverview,Shen2022DenoisingOverview}. Such degradations distort both spectral information and spatial structures, impairing the performance of subsequent tasks such as classification~\cite{Ghamisi2017Classification,Li2019Classification,Nicolas2019Classification} and unmixing~\cite{Bioucas-Dias2012UnmixingOverview,Ma2014UnmixingOverview}. Consequently, effective denoising is an essential preprocessing step for achieving high-precision HS image analysis.

To effectively remove noise from observed hyperspectral (HS) images, it is crucial to capture the intrinsic spatial–spectral properties of HS data and distinguish them from noise. While deep learning~\cite{Yuan2019HSIDCNN,Wei2021QRNN3D,Peng2024RCILD} and low-rank~\cite{Zhang2014LRMR, Xu2017LRMF, Zheng2020SpatDIs} approaches have been actively studied, they rely on training data or involve computationally expensive matrix/tensor decompositions, which limit their generalization or efficiency. In contrast, a total variation (TV)-based regularization approach has been developed in HS image denoising and destriping~\cite{ZHANG2024AEWTTV, Xu2024PWRCTV}, since it requires no training data and avoids costly decompositions. TV-based methods model the spatial–spectral piecewise smoothness inherent in HS images as the sparsity of spatial–spectral differences. In this work, we focus on the family of TV-based methods.

Existing TV-based methods for HS image denoising are largely extensions of the classical anisotropic and isotropic TV models originally developed for natural images~\cite{Rudin1992TV,Bresson2008TV}. One promising regularization method based on anisotropic TV is Spatio-Spectral Total Variation (SSTV)~\cite{Aggarwal2016SSTV}. SSTV is defined as computing the spatial differences after the spectral differences (referred to as the spatio-spectral second-order differences), and then aggregating them by the $\ell_{1}$-norm. Through this formulation, SSTV effectively captures the spatial and spectral piecewise smoothness. SSTV has been widely adopted as the foundation for state-of-the-art HS image denoising methods, such as incorporating an $\ell_{0}$-ball constraint~\cite{Ono2017l0Gradient} to strictly enforce sparsity in the spatial differences~\cite{Wang2021l0l1HTV}, constructing a spatial graph to better reflect complex spatial structures~\cite{Takemoto2022GSSTV}, and combining SSTV with low-rank approaches~\cite{Wang2018LRTDTV,Chen2023TPTV, Takemoto2025SSSTV}. However, since these methods are rooted in anisotropic TV that sums the vertical and horizontal differences separately, they measure oblique edges in a Manhattan-like manner, leading to the staircase effect. Moreover, these methods do not directly promote spatial smoothness, leaving artifacts.

In addition to the anisotropic formulation, Hybrid Spatio-Spectral Total Variation (HSSTV)~\cite{Takeyama2020HSSTV} has been proposed for HS image restoration in the isotropic formulation. In this formulation, HSSTV first groups spatial differences of each pixel with the $\ell_{2}$-norm and then aggregates them using the $\ell_{1}$-norm. This design preserves upward-sloping edges by evaluating them in the Euclidean manner. In natural image processing, this property makes isotropic TV more effective than anisotropic TV. Moreover, unlike SSTV, HSSTV evaluates not only spatio-spectral second-order differences but also spatial first-order differences, suppressing the artifacts that SSTV leaves. However, in the isotropic formulation, downward-sloping edges are still evaluated in the Manhattan-like manner (represented as the sum of the vertical and horizontal differences separately), making it difficult to accurately restore round structures. This raises the following research question: \textit{Can we design a TV-type regularization for HS image denoising that measures all edges and structures in the Euclidean manner to more accurately preserve round structures and oblique edges?}

In this paper, we propose a new regularization model, named Geometric Spatio-Spectral Total Variation (GeoSSTV), for HS image denoising. GeoSSTV is built upon the geometrically consistent framework proposed by Condat~\cite{Condat2017DTV}, which measures variations across any direction in the Euclidean manner. This geometric property is crucial for restoring round structures and oblique edges that are often corrupted by existing TV-based methods. The main contributions of this work are summarized as follows.
\begin{enumerate}
	\item We design a novel regularization formulation, namely GeoSSTV. GeoSSTV is designed to combine two types of TVs, consisting of the second-order spatio-spectral differences and the first-order spatial differences, thus suppressing artifacts that existing SSTV-type methods leave. Furthermore, since the two types of TVs are defined to be geometrically consistent formulations that measure variations across all directions in the Euclidean manner, GeoSSTV can preserve round structures and oblique edges while effectively removing noise.

	\item We formulate the mixed noise removal problem as a constrained convex optimization problem involving GeoSSTV. By explicitly characterizing Gaussian noise, sparse noise, and stripe noise with different convex constraints, the proposed method effectively removes the three types of noise. Moreover, modeling these terms as constraints rather than adding them to objective functions simplifies parameter tuning, as shown in prior studies~\cite{Afonso2011Constraint, Chierchia2015Constraint, Ono2015Constraint, Ono2017Constraint, Ono2019Constraint, Naganuma2022Destriping}.
	
	\item To solve the proposed optimization problem, we develop an efficient algorithm based on the preconditioned primal–dual splitting (P-PDS) method~\cite{Pock2011PPDS} with an operator norm-based step-size selection~\cite{Naganuma2023OVDP}. Unlike other popular algorithms used in existing HS image denoising methods, such as an alternating direction method of multipliers~\cite{Boyd2011ADMM} and PDS~\cite{Chambolle2011PDS, Condat2013PDS}, this approach automatically determines the stepsizes and guarantees stable convergence.
		
\end{enumerate}
Experimental results show the superiority of the proposed method to existing methods including state-of-the-art ones. The comparison of the features of the proposed and existing TV-based methods is summarized in Table~\ref{tab:ProsCons}. The preliminary version of this paper, without considering stripe noise, mathematical details, comprehensive experimental comparisons, or deeper discussions, has appeared in conference proceedings~\cite{Takemoto2024RISSTV}.
\begin{table*}[t]
    \begin{center}
        \caption{Pros and Cons of Existing and Proposed TV Methods for HS Image Denoising and Destriping.}
        \label{tab:ProsCons}
        \begin{tabular}{c ccccc}
            \toprule
                Methods & 
                \begin{tabular}{c}
                    Spatial-spectral \\ piecewise smoothness
                \end{tabular} & 
                Supressing staircase effect &
                Rotation invariance & 
                Supressing artifacts & 
                Convexity \\
            \cmidrule(lr){1-6}
            \vspace{-0.5mm}
                SSTV~\cite{Aggarwal2016SSTV} & $\checkmark$ & -- & $\checkmark$ & -- & $\checkmark$ \\
                $\llHTV$~\cite{Wang2021l0l1HTV} & $\checkmark$ & -- & $\checkmark$ & -- &-- \\
                GSSTV~\cite{Takemoto2022GSSTV} & $\checkmark$ & -- & $\checkmark$ & -- & $\checkmark$ \\
                LRTDTV~\cite{Wang2018LRTDTV} & $\checkmark$ & -- & $\checkmark$ & -- & -- \\
                TPTV~\cite{Chen2023TPTV} & $\checkmark$ & -- & $\checkmark$ & -- & -- \\
                HSSTV1~\cite{Takeyama2020HSSTV} & $\checkmark$ & -- & $\checkmark$ & $\checkmark$ & $\checkmark$ \\
                HSSTV2~\cite{Takeyama2020HSSTV} & $\checkmark$ & $\checkmark$ & -- & $\checkmark$ & $\checkmark$ \\
            \cmidrule(lr){1-6}
                Proposed method & $\checkmark$ & $\checkmark$ & $\checkmark$ & $\checkmark$ & $\checkmark$ \\
            \bottomrule
        \end{tabular}
    \end{center}
    \vspace{-3mm}
\end{table*}
\section{Preliminaries}
\label{sec:Preliminaries}

\subsection{Notations}
\label{subsec:Notations}
Throughout this paper, we denote vectors and matrices by boldface lowercase letters (e.g., $\NotaVector$) and boldface capital letters (e.g., $\NotaMatrix$), respectively. We consider an HS image, denoted by $\HSIClean$ with $\NumVert$ vertical pixels, $\NumHori$ horizontal pixels, and $\NumBand$ bands. We denote the total number of cube data elements by $\NumElement = \NumVert \NumHori \NumBand$. For matrix data $\NotaVector \in \RealSpace{\NumVert \NumHori}$, the value at location $(\IndexVert, \IndexHori)$ in the domain $\{1,\ldots,\NumVert\}\times\{1,\ldots,\NumHori\}$ is denoted by $[\NotaMatrix]_{\IndexVert,\IndexHori} \in \RealSpace{}$. 
A set of all proper lower semi-continuous convex functions over $\RealSpace{\NumVarOne}$ is denoted by $\SetProperLowerConvex{\RealSpace{\NumVarOne}}$.
The $\ell_{1}$-norm and the $\ell_{2}$-norm of a vector $\NotaVector \in \RealSpace{\NotaNumDim}$ are defined as $\| \NotaVector \|_{1} := \sum_{\NotaIndex=1}^{\NotaNumDim} | \NotaScalar_{\NotaIndex} |$ and $\| \NotaVector \|_{2} := \sqrt{\sum_{\NotaIndex=1}^{\NotaNumDim} \NotaScalar_{\NotaIndex}^{2}}$, respectively, where  $\NotaScalar_{\NotaIndex}$ represents the $\NotaIndex$-th entry of $\NotaVector$. 
For an HS image $\HSIClean \in \RealSpace{\NumElement}$, let $\DiffOpVert \in \RealSpace{\NumElement \times \NumElement}$, $\DiffOpHori \in \RealSpace{\NumElement \times \NumElement}$, and $\DiffOpBand \in \RealSpace{\NumElement \times \NumElement}$ be the forward difference operators in the horizontal, vertical, and spectral directions, respectively, and the boundary condition is the Neumann boundary. Here, spatial difference operator is denoted by $\DiffOpSp := \begin{pmatrix} \DiffOpVertT & \DiffOpHoriT \end{pmatrix}^{\top} \in \RealSpace{2 \NumElement \times \NumElement}$. Other notations will be introduced as needed.

\subsection{Proximal Tools}
\label{subsec:Prox}
In this chapter, we introduce basic proximal tools that play a central role in the optimization part of our method.
For any $\CoefProx > 0$, the proximity operator of $\FuncProx \in \SetProperLowerConvex{\RealSpace{\NumVarProx}}$ is defined by
\begin{equation}
	\label{eq:Prox}
	\prox_{\CoefProx \FuncProx}(\VarProxOne) := \argmin_{\VarProxTwo \in \RealSpace{\NumVarProx}} \FuncProx(\VarProxTwo) + \frac{1}{2 \CoefProx} \| \VarProxOne - \VarProxTwo \|_{2}^{2}.
\end{equation}

The \textit{Fenchel--Rockafellar conjugate function} $\FuncProx^{*}$ of the function $\FuncProx \in \SetProperLowerConvex{\RealSpace{\NumVarProx}}$ is defined by
\begin{equation}
	\label{eq:ConjugateFunction}
	\FuncProx^{*}(\VarProxOne) := \sup_{\VarProxTwo} \InnerProduct<\VarProxOne, \VarProxTwo> - \FuncProx(\VarTwo),
\end{equation}
where $\InnerProduct<\cdot , \cdot>$ is the Euclidean inner product.
Thanks to the generalization of Moreau's identity~\cite{Combettes2013Moreau}, the proximity operator of $\FuncProx^{*}$ is calculated as
\begin{equation}
	\label{eq:ProxConjugate}
	\prox_{\CoefProx \FuncProx^{*}}(\VarProxOne) = \VarProxOne - \CoefProx \prox_{\frac{1}{\CoefProx} \FuncProx} \left( \frac{1}{\CoefProx} \VarProxOne \right).
\end{equation}

The indicator function $\FuncIndicator{\SetConvex}$ of a nonempty closed convex set $\SetConvex \subset \RealSpace{\NumVarOne}$ belongs to $\SetProperLowerConvex{\RealSpace{\NumVarOne}}$ and is defined as 
\begin{equation}
	\label{eq:Indicator_Function}
	\FuncIndicator{\SetConvex} (\VarOne) := 
	\begin{cases}
		0, & \mathrm{if} \: \VarOne \in \SetConvex, \\
		\infty, & \mathrm{otherwise}.
	\end{cases}
\end{equation}
The proximity operator of $\FuncIndicator{\SetConvex}$ is equivalent to the projection onto $\SetConvex$, as given by
\begin{equation}
	\label{eq:Projection}
	\prox_{\CoefProx \FuncIndicator{\SetConvex}} (\VarProxOne) = \Projection{\SetConvex}(\VarProxOne) := \argmin_{\VarProxTwo \in \SetConvex} \|\VarProxTwo - \VarProxOne \|_{2}.
\end{equation}

\subsection{Preconditoned Primal-Dual Splitting Method (P-PDS)}
\label{subsec:P-PDS}
P-PDS~\cite{Pock2011PPDS}, on which our algorithm is based, solves the following generic form of convex optimization problems:
\begin{align}
	\label{prob:convex_optim_prob}
	\min_{\substack{\VarPrimal{1}, \ldots, \VarPrimal{\NumVarPrimal}, \\ 
			\VarDual{1}, \ldots, \VarDual{\NumVarDual}}} 
	& \sum_{\IndexPrimal=1}^{\NumVarPrimal} \FuncPrimal{\IndexPrimal} (\VarPrimal{\IndexPrimal}) + \sum_{\IndexDual=1}^{\NumVarDual} \FuncDual{\IndexDual} (\VarDual{\IndexDual}) \nonumber \\ 
	& \mathrm{s.t.} \:
	\begin{cases} 
		\VarDual{1} = \sum_{\IndexPrimal=1}^{\NumVarPrimal} \LinOpPPDS{1,\IndexPrimal} \VarPrimal{\IndexPrimal}, \\ 
		\vdots \\ 
		\VarDual{\NumVarDual} = \sum_{\IndexPrimal=1}^{\NumVarPrimal} \LinOpPPDS{\NumVarDual,\IndexPrimal} \VarPrimal{\IndexPrimal}, 
	\end{cases}
\end{align}
where $\FuncPrimal{\IndexPrimal} \in \SetProperLowerConvex{\RealSpace{\DimVarPrimal{\IndexPrimal}}} \: (\IndexPrimal = 1, \ldots, \NumVarPrimal)$ and $\FuncDual{\IndexDual} \in \SetProperLowerConvex{\RealSpace{\DimVarDual{\IndexDual}}} \: (\IndexDual = 1, \ldots, \NumVarDual)$, $\VarPrimal{\IndexPrimal} \in \RealSpace{\DimVarPrimal{\IndexPrimal}} \: (\IndexPrimal = 1, \dots, \NumVarPrimal)$ are primal variables, $\VarDual{\IndexDual} \in \RealSpace{\DimVarDual{\IndexDual}} \: (\IndexDual = 1, \dots, \NumVarDual)$ are dual variables, and $\LinOpPPDS{\IndexDual,\IndexPrimal} \in \RealSpace{\DimVarDual{\IndexDual} \times  \DimVarPrimal{\IndexPrimal}}$ ($i = 1, \ldots, \NumVarPrimal$, $j = 1, \ldots, \NumVarDual$) are linear operators.

Using the proximity operators of the functions in Prob.~\eqref{prob:convex_optim_prob}, P-PDS is given by the following iterative procedures:
\begin{equation}
	\label{algo:P-PDS}
	\AlgoLfloor{
		\begin{array}{l}
			\VarPrimal{1}^{(\IndexAlg+1)} 
			\leftarrow \prox_{\ScalarStepsize{1, 1} \FuncPrimal{1}}
			\left( \VarPrimal{1}^{(\IndexAlg)} - \ScalarStepsize{1,1} \bigl( \sum_{\IndexDual=1}^{\NumVarDual} \LinOpPPDS{\IndexDual,1}^{\top} \VarDual{\IndexDual}^{(\IndexAlg)} \bigr) \right), \\
			\vdots \\
			\VarPrimal{\NumVarPrimal}^{(\IndexAlg+1)} 
			\leftarrow \prox_{\ScalarStepsize{1, \NumVarPrimal} \FuncPrimal{\NumVarPrimal}}
			\left( \VarPrimal{\NumVarPrimal}^{(\IndexAlg)} - \ScalarStepsize{1, \NumVarPrimal} \bigl( \sum_{\IndexDual=1}^{\NumVarDual} \LinOpPPDS{\IndexDual,\NumVarPrimal}^{\top} \VarDual{\IndexDual}^{(\IndexAlg)} \bigr) \right), \\
			\VarPrimal{\IndexPrimal}^{'} = 2\VarPrimal{\IndexPrimal}^{(\IndexAlg+1)} - \VarPrimal{\IndexPrimal}^{(\IndexAlg)} \: (\forall \IndexPrimal = 1, \ldots, \NumVarPrimal), \\
			\VarDual{1}^{(\IndexAlg+1)}
			\leftarrow \prox_{\ScalarStepsize{2, 1} \FuncDual{1}^{*}}
			\left( \VarDual{1}^{(\IndexAlg)} - \ScalarStepsize{2, 1}
			\bigl(\sum_{\IndexPrimal=1}^{\NumVarPrimal} \LinOpPPDS{1,\IndexPrimal} \VarPrimal{\IndexPrimal}^{'} \bigr) \right), \\
			\vdots \\
			\VarDual{\NumVarDual}^{(\IndexAlg+1)}
			\leftarrow \prox_{\ScalarStepsize{2, \NumVarDual} \FuncDual{\NumVarDual}^{*}}
			\left( \VarDual{\NumVarDual}^{(\IndexAlg)} - \ScalarStepsize{2, \NumVarDual}
			\bigl(\sum_{\IndexPrimal=1}^{\NumVarPrimal} \LinOpPPDS{\NumVarDual,\IndexPrimal}  \VarPrimal{\IndexPrimal}^{'} \bigr) \right), \\
		\end{array}
	}.
\end{equation}
where $\ScalarStepsize{1, \IndexPrimal}(\IndexPrimal = 1, \dots, \NumVarPrimal)$ and $\ScalarStepsize{2, \IndexDual}(\IndexDual = 1, \dots, \NumVarDual)$ are the stepsize parameters.

In contrast to the standard PDS~\cite{Chambolle2011PDS, Condat2013PDS}, P-PDS can automatically determine the stepsize parameters as follows~\cite{Naganuma2023OVDP}:
\begin{equation} 
	\label{eq:stepsize}
	\ScalarStepsize{1, \IndexPrimal} = \frac{1}{\sum_{\IndexDual = 1}^{\NumVarDual} \OpNormSq{ \LinOpPPDS{\IndexDual, \IndexPrimal} }}, \quad \ScalarStepsize{2, \IndexDual} = \frac{1}{\NumVarPrimal},
\end{equation}
where $\OpNorm{\cdot}$ is the operator norm defined by
\begin{equation}
	\label{eq:opnorm}
	\OpNorm{\LinOpPPDS{}} := \sup_{\VarPrimal{} \neq \mathbf{0}} \frac{\| \LinOpPPDS{} \VarPrimal{} \|_{2}}{\| \VarPrimal{} \|_{2}}.
\end{equation}

\subsection{Condat's geometrically consistent TV~\cite{Condat2017DTV}}
\label{subsec:DTV}
Condat proposed a new total variation (TV) formulation that measures variations across all directions in a Euclidean manner for grayscale images~\cite{Condat2017DTV}.For given grayscale image be $\VarCTVImage \in \RealSpace{\NumVert \NumHori}$, it is defined in the dual domain as follows: for all $\IndexVert = 1,\ldots,\NumVert$, $\IndexHori = 1,\ldots,\NumHori$,
\begin{equation}
	\label{eq:DTVDual}
	\CondatTV (\VarCTVImage) := \max_{\VarCTVDual \in \RealSpace{2\NumVert \NumHori}} \InnerProduct<\DiffOpSpCTV \VarCTVImage, \VarCTVDual> \: \mathrm{s.t.} \: 
	\begin{cases}
		\| \OpIpoVert \VarCTVDual \|_{1, \infty} \leq 1, \\
		\| \OpIpoHori \VarCTVDual \|_{1, \infty} \leq 1, \\
		\| \OpIpoCent \VarCTVDual \|_{1, \infty} \leq 1,
	\end{cases}
\end{equation}
where $\DiffOpSpCTV \in \RealSpace{2\NumVert \NumHori \times \NumVert \NumHori}$ is the spatial difference operator for the grayscale image, $\VarCTVDual = (\VarCTVDualVert^{\top}, \VarCTVDualHori^{\top})^{\top} \in \RealSpace{2\NumVert \NumHori}$ is the dual image pair, and $\| \cdot \|_{1, \infty} := \max_{\IndexVert, \IndexHori} \| [\cdot]_{\IndexVert, \IndexHori} \|_{1}$.
The dual images $[\VarCTVDualVert]_{\IndexVert, \IndexHori}$ and $[\VarCTVDualHori]_{\IndexVert, \IndexHori}$, corresponding to $[\DiffOpVert \VarCTVImage]_{\IndexVert, \IndexHori}$ and $[\DiffOpHori \VarCTVImage]_{\IndexVert, \IndexHori}$, are naturally located at the half-pixel shifted positions $(\IndexVert+\tfrac{1}{2},\IndexHori)$ and $(\IndexVert,\IndexHori+\tfrac{1}{2})$, respectively. To avoid the inconsistency caused by this half-pixel shift, Condat introduced interpolation operators $\OpIpoVert$, $\OpIpoHori$, $\OpIpoCent \in \RealSpace{2\NumVert \NumHori \times 2\NumVert \NumHori}$ (see Eqs.~(9)–(14) in \cite{Condat2017DTV}), which correct these shifts and realign the dual variables onto the grids $(\IndexVert+\tfrac{1}{2},\IndexHori)$, $(\IndexVert,\IndexHori+\tfrac{1}{2})$, and $(\IndexVert,\IndexHori)$, respectively. This design enables grouping without directional bias, so that variations across all directions are measured in the Euclidean manner, resulting in a geometrically consistent formulation of TV.
The dual formulation Eq.~\eqref{eq:DTVDual} can be rewritten into the equivalent primal formulation as
\begin{align}
	\label{eq:DTVPrimal}
	\CondatTV (\VarCTVImage) = &\min_{\VarCTVGradVert, \VarCTVGradHori, \VarCTVGradCent \in \RealSpace{2\NumVert \NumHori}} \|\VarCTVGradVert \|_{1,2} + \|\VarCTVGradHori \|_{1,2} + \|\VarCTVGradCent \|_{1,2} \nonumber \\
	&\mathrm{s.t.} \: \OpIpoVertT \VarCTVGradVert + \OpIpoHoriT \VarCTVGradHori + \OpIpoCentT \VarCTVGradCent = \DiffOpSpCTV \VarCTVImage,
\end{align}
where $\| \cdot \|_{1,2}$ is the mixed $\ell_{1,2}$ norm grouping the vertical and horizontal directions.
The three vectors $\VarCTVGradVert$, $\VarCTVGradHori$, $\VarCTVGradCent$ are viewed as gradients on the grids $(\IndexVert+\tfrac{1}{2}, \IndexHori)$, $(\IndexVert, \IndexHori+\tfrac{1}{2})$, $(\IndexVert, \IndexHori)$, respectively.
For a more compact form, let the linear operator $\OpIpoCTV = \begin{pmatrix}\OpIpoVertT & \OpIpoHoriT & \OpIpoCentT \end{pmatrix}^{\top} \in \RealSpace{6\NumVert \NumHori \times 2\NumVert \NumHori}$, and the vector $\VarCTVGrad = \begin{pmatrix} \VarCTVGradVertT &\VarCTVGradHoriT & \VarCTVGradCentT \end{pmatrix}^{\top}\in \RealSpace{6\NumVert \NumHori}$.
Then Eq.~\eqref{eq:DTVPrimal} can be written as
\begin{equation}
	\label{eq:DTVPrimalCmpct}
	\CondatTV (\VarCTVImage) = \min_{\VarCTVGrad \in \RealSpace{6\NumVert \NumHori}}  \|\VarCTVGrad \|_{1,2} \quad \mathrm{s.t.} \: \OpIpoCTVT \VarCTVGrad = \DiffOpSpCTV \VarCTVImage. 
\end{equation}
\section{Proposed Method}
\label{sec:Proposed}

\subsection{Geometric Spatio-Spectral Total Variation}
\label{subsec:GeoSSTV}
Combining the second-order spatio-spectral differences with the first-order spatial differences and having a geometrically consistent property that evaluates variations in the Euclidean manner across all directions, our GeoSSTV is defined as follows:
\begin{align}
	\label{eq:GeoSSTV}
	\GeoSSTV (\HSIClean) := &\min_{\IpoOne, \IpoTwo} \ParamBalance \| \IpoOne \|_{1,2} + \| \IpoTwo \|_{1,2} \nonumber \\
	&\mathrm{s.t.} \: 
	\begin{cases}
		\OpIpoT \IpoOne = \DiffOpSp \HSIClean, \\
		\OpIpoT \IpoTwo = \DiffOpSp \DiffOpBand \HSIClean,
	\end{cases}
\end{align}
where $\ParamBalance \geq 0$, $\IpoOne$, $\IpoTwo \in \RealSpace{6\NumElement}$ are auxiliary variables, $\OpIpo \in 2 \NumElement \times 6\NumElement$ is linear operator formed by arranging $\NumBand$ diagonals of $\OpIpoCTV$ defined in Eq.~\eqref{eq:DTVPrimalCmpct}. Mainly, the second-order spatio-spectral TV corresponding to the second term and the second constraint in Eq.~\eqref{eq:GeoSSTV} characterizes the spatio-spectral piecewise smoothness inherent in HS images. Associated with the first term and the first constraint, the first-order spatial TV plays an supplementary role to suppress the noise-like artifacts produced by only imposing the second-order TV. These two parts are formulated under the geometrically consistent framework, thus GeoSSTV achieves round structures and oblique edges in HS images more accurately than existing TV-based methods.

The parameter $\ParamBalance$ controls the relative importance of the two types of TV.
If $\ParamBalance$ is larger, i.e., we make the direct spatial smoothness stronger on a restored HS image, GeoSSTV would cause over-smoothing of the detailed structures.
Therefore, $\ParamBalance$ should be set to less than one.

\subsection{HS Image Denoising by GeoSSTV}
\label{subsec:HSIDenoising}
An observed HS image $\HSIObsv \in \RealSpace{\NumElement}$ contaminated by mixed noise is modeled by
\begin{equation}
	\label{eq:ObservationModel}
	\HSIObsv = \bar{\HSIClean} + \bar{\NoiseSparse} + \bar{\NoiseStripe} + \NoiseGauss,
\end{equation}
where $\bar{\HSIClean}$ is a clean HS image, $\bar{\NoiseSparse}$ is sparse noise that models outliers and deadline noise, $\bar{\NoiseStripe}$ is stripe noise, and $\NoiseGauss$ is Gaussian noise that models random noise, respectively.

Based on the above observation model, we formulate the HS image denoising problem that handles GeoSSTV as a constrained convex optimization problem with the following form:
\begin{equation}
	\label{prob:GeoSSTV_denoising}
	\min_{\HSIClean, \IpoOne, \IpoTwo, \NoiseSparse, \NoiseStripe} \ParamBalance \| \IpoOne \|_{1,2} + \| \IpoTwo \|_{1,2} \: \mathrm{s.t.}
	\begin{cases}
		\NoiseSparse \in \BallSparse, \\
		\NoiseStripe \in \BallStripe, \\
		\DiffOpVert \NoiseStripe = \mathbf{0}, \\
		\HSIClean + \NoiseSparse + \NoiseStripe \in \BallFidel, \\
		\HSIClean \in \SetRange, \\
		\OpIpoT \IpoOne = \DiffOpSp \HSIClean, \\
		\OpIpoT \IpoTwo = \DiffOpSp \DiffOpBand \HSIClean,
	\end{cases}
\end{equation}
where
\begin{align}
	\label{eq:constraint_sparse}
	&\BallSparse := \{ \VarOne \in \RealSpace{\NumElement} | \:
	\|\VarOne\|_{1} \leq \RadiusSparse \}, \\
	\label{eq:constraint_stripe}
	&\BallStripe := \{ \VarOne \in \RealSpace{\NumElement} | \:
	\|\VarOne\|_{1} \leq \RadiusStripe \},  \\
	\label{eq:constraint_fidel}
	&\BallFidel := \{ \VarOne \in \RealSpace{\NumElement} | \:
	\|\VarOne - \HSIObsv\|_2 \leq \RadiusFidel \}, \\
	\label{eq:constraint_box}
	&\SetRange := \{ \VarOne \in \RealSpace{\NumElement} | \:
	\MinRange \leq \ElementOne{\IndexOne} \leq \MaxRange  \: (\IndexOne = 1,\dots , \NumElement) \}.
\end{align}
The first constraint characterizes sparse noise $\NoiseSparse$ with the zero-centered $\ell_1$-ball of the radius $\RadiusSparse > 0$. 
The second constraint controls the intensity of stripe noise $\NoiseStripe$ and the third constraint captures the vertical flatness property by imposing zero to the vertical gradient of $\NoiseStripe$. These constraints effectively characterize stripe noise~\cite{Naganuma2021Flatness}.
The fourth constraint serves as data-fidelity with the $\HSIObsv$-centered $\ell_2$-ball of the radius $\RadiusFidel > 0$.
The fifth constraint is a box constraint with $\MinRange < \MaxRange$ which represents the dynamic range of $\HSIClean$.
For HS images where each element is normalized, we can set $\MinRange = 0$ and $\MaxRange = 1$.
We impose the first, second, and fourth constraints instead of adding terms to the objective function. This formulation allows the hyperparameters $\RadiusSparse$, $\RadiusStripe$, and $\RadiusFidel$ to be determined independently, making it easier to adjust them.
The effectiveness of such constraint-based modeling has been discussed in~\cite{Afonso2011Constraint, Chierchia2015Constraint, Ono2015Constraint, Ono2017Constraint, Ono2019Constraint, Naganuma2022Destriping}.

Using indicator functions $\FuncIndicator{\SetZero}$, $\FuncIndicator{\BallFidel}$, $\FuncIndicator{\BallStripe}$, $\FuncIndicator{\BallSparse}$, and $\FuncIndicator{\SetRange}$, we rewrite Prob.~\eqref{prob:GeoSSTV_denoising} into an equivalent form:
\begin{align}
	\label{prob:denoising2PPDS}
	\min_{
		\substack{
			\HSIClean, \IpoOne, \IpoTwo, \NoiseSparse, \NoiseStripe, \\ 
			\VarDual{1}, \VarDual{2}, \VarDual{3}}} \:
	& \ParamBalance \| \IpoOne \|_{1,2} + \| \IpoTwo \|_{1,2} \notag \\
	& + \FuncIndicator{\SetRange} (\HSIClean) + \FuncIndicator{\BallSparse} (\NoiseSparse) + \FuncIndicator{\BallStripe} (\NoiseStripe) \notag \\
	& + \FuncIndicator{\SetZero} (\VarDual{1}) + \FuncIndicator{\SetZero} (\VarDual{2}) + \FuncIndicator{\SetZero} (\VarDual{3}) + \FuncIndicator{\BallFidel} (\VarDual{4}) \nonumber \\
	& \mathrm{s.t.} \:
	\begin{cases} 
		\VarDual{1} = \DiffOpSp \HSIClean - \OpIpoT \IpoOne, \\
		\VarDual{2} = \DiffOpSp \DiffOpBand \HSIClean - \OpIpoT \IpoTwo, \\
		\VarDual{3} = \DiffOpVert \NoiseStripe, \\
		\VarDual{4} = \HSIClean + \NoiseSparse + \NoiseStripe.
	\end{cases}
\end{align}
Prob.~\eqref{prob:denoising2PPDS} can be solved by P-PDS~\cite{Pock2011PPDS}.
We show the detailed algorithm in Alg.~1.
The proximity operators of $\FuncIndicator{\SetRange}$, $\FuncIndicator{\SetZero}$, $\FuncIndicator{\BallFidel}$, and $\| \cdot \|_{1,2}$ are calculated by
\begin{align}
	\label{eq:prox_box_constraint}
	&[\prox_{\ParamStepsize{} \FuncIndicator{\SetRange}} (\VarOne)]_{\IndexProx} 
	 = [\Projection{\SetRange} (\VarOne)]_{\IndexProx} =
	\begin{cases} 
		\MinRange, & \text{if } \ElementOne{\IndexProx} < \MinRange, \\ 
		\MaxRange, & \text{if } \ElementOne{\IndexProx}> \MaxRange, \\ 
		\ElementOne{\IndexProx}, & \text{otherwise,}
	\end{cases} \\
	\label{eq:prox_zeroset}
	&\prox_{\ParamStepsize{} \FuncIndicator{\SetZero}}(\VarOne) = \mathbf{0}, \\
	\label{eq:prox_l2ball_constraint}
	&\prox_{\ParamStepsize{} \FuncIndicator{\BallFidel}}(\VarOne)
	= \Projection{\BallFidel} (\VarOne) = 
	\begin{cases}
		\VarOne, & \text{if } \VarOne \in \BallFidel, \\ 
		\HSIObsv + \frac{\varepsilon (\VarOne - \HSIObsv)}{\| \VarOne - \HSIObsv \|_2}, & \text{otherwise,}
	\end{cases} \\
	\label{eq:prox_l12norm}
	&\prox_{\ParamStepsize{} \| \cdot \|_{1,2}}(\VarOne_{\GroupIndex})
	= \max\left\{1 - \frac{\ParamStepsize{}}{\|\VarOne_{\GroupIndex}\|_2}, 0\right\} \VarOne_{\GroupIndex},
\end{align}
where $\VarOne_{\GroupIndex} \in \RealSpace{2}$ denotes the subvector of $\VarOne$ corresponding to group $\GroupIndex$, consisting of a pair of vertical and horizontal differences at each pixel.
The proximity operators of $\FuncIndicator{\BallSparse} (\NoiseSparse)$ can be efficiently computed by a fast $\ell_{1}$-ball projection algorithm~\cite{Condat2016L1ball}.

Based on Eq.~\eqref{eq:stepsize}, we set the stepsize parameters $\ParamStepsize{\HSIClean} = \tfrac{1}{13}$, $\ParamStepsize{\IpoOne} = \tfrac{1}{4}$, $\ParamStepsize{\IpoTwo} = \tfrac{1}{4}$, $\ParamStepsize{\NoiseSparse} = 1$, $\ParamStepsize{\NoiseStripe} = \tfrac{1}{3}$, $\ParamStepsize{\VarDual{1}} = \ParamStepsize{\VarDual{2}} = \ParamStepsize{\VarDual{3}} = \ParamStepsize{\VarDual{4}} = \tfrac{1}{5}$.

\begin{figure}[!t]
	\begin{algorithm}[H]
	    \caption{P-PDS-based solver for \eqref{prob:denoising2PPDS}}
		\label{algo_PPDS_for_denoising}
		\begin{algorithmic}[1]
			\renewcommand{\algorithmicrequire}{\textbf{Input:}}
			\renewcommand{\algorithmicensure}{\textbf{Output:}}
			\REQUIRE $\HSIClean^{(0)}, \NoiseSparse^{(0)}, \NoiseStripe^{(0)}, \IpoOne^{(0)},  \IpoTwo^{(0)}, \VarDual{1}^{(0)}, \VarDual{2}^{(0)}, \VarDual{3}^{(0)}$
			\ENSURE $\HSIClean^{(\IndexAlg)}$
			\WHILE {A stopping criterion is not satisfied}
                    \STATE $\HSIClean^{(\IndexAlg+1)} \leftarrow$ \\ $\Projection{\SetRange}\left( \HSIClean^{(\IndexAlg)} - \ParamStepsize{\HSIClean} \bigl(\DiffOpSpT \VarDual{1}^{(\IndexAlg)} + \DiffOpBandT \DiffOpSpT \VarDual{2}^{(\IndexAlg)} + \VarDual{4}^{(\IndexAlg)} \bigr) \right)$
                    \STATE $\NoiseSparse^{(\IndexAlg + 1)} \leftarrow \Projection{\BallSparse} \left( \NoiseSparse^{(\IndexAlg)} - \ParamStepsize{\NoiseSparse} \VarDual{4}^{(\IndexAlg)} \right)$
                    \STATE $\NoiseStripe^{(\IndexAlg + 1)} \leftarrow \Projection{\BallStripe} \left( \NoiseStripe^{(\IndexAlg)} - \ParamStepsize{\NoiseStripe} \bigl( \DiffOpVertT \VarDual{3}^{(\IndexAlg)} + \VarDual{4}^{(\IndexAlg)} \bigr) \right)$
                    \STATE $\IpoOne^{(\IndexAlg + 1)} \leftarrow \prox_{\ParamStepsize{\IpoOne}\ParamBalance \| \cdot \|_{1,2}} \left( \IpoOne^{(\IndexAlg)} + \ParamStepsize{\IpoOne} \OpIpo \VarDual{1}^{(\IndexAlg)} \right)$
                    \STATE $\IpoTwo^{(\IndexAlg + 1)} \leftarrow \prox_{\ParamStepsize{\IpoTwo} \| \cdot \|_{1,2}} \left( \IpoTwo^{(\IndexAlg)} + \ParamStepsize{\IpoTwo} \OpIpo \VarDual{2}^{(\IndexAlg)} \right)$
                    \STATE $\HSICleanRes \leftarrow 2 \HSIClean^{(\IndexAlg+1)} - \HSIClean^{(\IndexAlg)}$;
                    \STATE $\ResNoiseSparse \leftarrow 2 \NoiseSparse^{(\IndexAlg+1)} - \NoiseSparse^{(\IndexAlg)}$;
                    \STATE $\ResNoiseStripe \leftarrow 2 \NoiseStripe^{(\IndexAlg+1)} - \NoiseStripe^{(\IndexAlg)}$;
                    \STATE $\IpoOneRes \leftarrow 2 \IpoOne^{(\IndexAlg+1)} - \IpoOne^{(\IndexAlg)}$;
                    \STATE $\IpoTwoRes \leftarrow 2 \IpoTwo^{(\IndexAlg+1)} - \IpoTwo^{(\IndexAlg)}$;
                    \STATE $\VarDual{1}^{(\IndexAlg+1)} \leftarrow \VarDual{1}^{(\IndexAlg)} + \ParamStepsize{\VarDual{1}} \left( \DiffOpSp \HSICleanRes - \OpIpoT \IpoOneRes  \right)$
                    \STATE $\VarDual{2}^{(\IndexAlg+1)} \leftarrow \VarDual{2}^{(\IndexAlg)} + \ParamStepsize{\VarDual{2}} \left( \DiffOpSp \DiffOpBand \HSICleanRes - \OpIpoT \IpoTwoRes \right)$
                    \STATE $\VarDual{3}^{(\IndexAlg+1)} \leftarrow \VarDual{3}^{(\IndexAlg)} + \ParamStepsize{\VarDual{3}} \DiffOpVert \ResNoiseStripe$ 
                    \STATE $\VarDual{4}^{'} \leftarrow \VarDual{4}^{(\IndexAlg)} + \ParamStepsize{\VarDual{4}} ( \HSICleanRes + \ResNoiseSparse + \ResNoiseStripe )$;
                    \STATE $\VarDual{4}^{(\IndexAlg+1)} \leftarrow \VarDual{4}^{'} - \ParamStepsize{\VarDual{4}} \Projection{\BallFidel} \left(\frac{1}{\ParamStepsize{\VarDual{4}}} \VarDual{4}^{'} \right)$;
    			\STATE $\IndexAlg \leftarrow \IndexAlg + 1$;
			\ENDWHILE
		\end{algorithmic}
	\end{algorithm}
	\vspace{-2mm}
\end{figure}
\section{Experiments}
\label{sec:experiments}
\begin{table*}[t]
    \renewcommand{\arraystretch}{0.75}
    \begin{center}
        \caption{MPSNRs and MSSIMs of All Denoising Results in simulated experiments.}
        \label{tab:MPSNR_MSSIM}
        		\scalebox{0.90}{
        \begin{tabular}{ccc ccccccc}
            \toprule
                Image & Case & Metric & SSTV~\cite{Aggarwal2016SSTV} & $\llHTV$~\cite{Wang2021l0l1HTV} & HSSTV1~\cite{Takeyama2020HSSTV} & HSSTV2~\cite{Takeyama2020HSSTV} & TPTV~\cite{Chen2023TPTV} & QRNN3D~\cite{Wei2021QRNN3D} & GeoSSTV \\
            \cmidrule(lr){1-10} 
            \multirow{10}{*}{Pavia University} 
            & \multirow{2}{*}{Case 1} & 
            MPSNR & 35.96 & 35.86 & 36.44 & \underline{36.65} & 33.99 & 35.18 & \textbf{36.81} \\ 
            & & MSSIM & 0.9231 & 0.9207 & 0.9388 & \underline{0.9418} & 0.8850 & 0.9319 & \textbf{0.9420} \\ 
            \cmidrule(lr){2-10} 
            & \multirow{2}{*}{Case 2} & 
            MPSNR & 33.04 & 33.45 & 33.62 & 33.63 & \underline{33.68} & 32.57 & \textbf{34.24} \\ 
            & & MSSIM & 0.8763 & 0.8908 & 0.8895 & 0.8865 & 0.8795 & \underline{0.8975} & \textbf{0.9079} \\ 
            \cmidrule(lr){2-10} 
            & \multirow{2}{*}{Case 3} & 
            MPSNR & 34.14 & 33.96 & 34.98 & \underline{35.14} & 33.46 & 32.51 & \textbf{35.21} \\ 
            & & MSSIM & 0.8871 & 0.8832 & 0.9123 & \underline{0.9151} & 0.8795 & 0.9011 & \textbf{0.9179} \\ 
            \cmidrule(lr){2-10} 
            & \multirow{2}{*}{Case 4} & 
            MPSNR & 34.76 & 34.90 & 34.66 & 34.79 & 33.93 & \underline{35.23} & \textbf{35.58} \\ 
            & & MSSIM & 0.9142 & 0.9167 & 0.9207 & 0.9217 & 0.8901 & \underline{0.9316} & \textbf{0.9356} \\ 
            \cmidrule(lr){2-10} 
            & \multirow{2}{*}{Case 5} & 
            MPSNR & 34.90 & 34.60 & 34.96 & \underline{35.10} & 33.22 & 34.29 & \textbf{35.55} \\ 
            & & MSSIM & 0.9117 & 0.9048 & 0.9237 & \underline{0.9256} & 0.8733 & 0.9211 & \textbf{0.9289} \\ 

            \cmidrule(lr){1-10} 
            \multirow{10}{*}{Jasper Ridge} 
            & \multirow{2}{*}{Case 1} & 
            MPSNR & 35.97 & 35.73 & 36.24 & \underline{36.33} & 36.21 & 32.69 & \textbf{36.45} \\ 
            & & MSSIM & 0.9168 & 0.9115 & 0.9365 & \underline{0.9386} & 0.9279 & 0.8761 & \textbf{0.9394} \\ 
            \cmidrule(lr){2-10} 
            & \multirow{2}{*}{Case 2} & 
            MPSNR & 34.35 & \underline{34.79} & 34.15 & 34.09 & 34.31 & 32.22 & \textbf{34.86} \\ 
            & & MSSIM & 0.9007 & \underline{0.9157} & 0.9097 & 0.9069 & 0.8789 & 0.8673 & \textbf{0.9203} \\ 
            \cmidrule(lr){2-10} 
            & \multirow{2}{*}{Case 3} & 
            MPSNR & 34.08 & 33.88 & 34.79 & \underline{34.87} & 32.71 & 32.44 & \textbf{35.01} \\ 
            & & MSSIM & 0.8770 & 0.8732 & 0.9046 & \underline{0.9062} & 0.8285 & 0.8734 & \textbf{0.9106} \\ 
            \cmidrule(lr){2-10} 
            & \multirow{2}{*}{Case 4} & 
            MPSNR & 34.83 & 34.87 & 34.19 & 34.23 & \underline{35.02} & 32.79 & \textbf{35.12} \\ 
            & & MSSIM & 0.9167 & 0.9139 & 0.9203 & 0.9199 & \underline{0.9217} & 0.8774 & \textbf{0.9310} \\ 
            \cmidrule(lr){2-10} 
            & \multirow{2}{*}{Case 5} & 
            MPSNR & \underline{34.83} & 34.55 & 34.33 & 34.33 & 34.09 & 32.07 & \textbf{35.18} \\ 
            & & MSSIM & 0.9077 & 0.8987 & \underline{0.9172} & 0.9169 & 0.9076 & 0.8660 & \textbf{0.9268} \\ 

            \bottomrule

        \end{tabular}
        		}
    \end{center}
    \vspace{-1mm}
\end{table*}
To demonstrate the effectiveness of GeoSSTV, we conducted mixed noise removal experiments on HS image contaminated with simulated or real noise.
We compared GeoSSTV with four types of methods; SSTV-based methods, i.e., SSTV~\cite{Aggarwal2016SSTV}, $\llHTV$~\cite{Wang2021l0l1HTV}, and HSSTV~\cite{Takeyama2020HSSTV}; TV-LR hybrid method, i.e., TPTV~\cite{Chen2023TPTV}; and DNN-based method, i.e., QRNN3D~\cite{Wei2021QRNN3D}. 
Here, HSSTV with $\ell_{1}$-norm and $\ell_{1,2}$-norm are denoted by HSSTV1 and HSSTV2, respectively.
For a fair comparison, the regularization functions of the P-PDS applicable methods, i.e., SSTV, HSSTV1, HSSTV2, and $\llHTV$ were replaced with the GeoSSTV regularization function in Prob.~\eqref{prob:GeoSSTV_denoising}, and we solve each problem by P-PDS.
For TPTV and QRNN3D, we used implementation codes published by the authors\footnote{The TPTV and QRNN3D implementation codes are available at
\\ https://github.com/chuchulyf/ETPTV \\ and https://github.com/Vandermode/QRNN3D?tab=readme-ov-file.}. For QRNN3D, we performed fine-tuning using Pavia Centre\footnote{\url{https://www.ehu/ccwintco/index/php/Hyperspectral_Remote_Sensing_Scenes}}\label{fn:PaviaURL}to improve noise removal performance.

\subsection{Simulated HS Image Experiments}
\begin{figure*}[t]
	\begin{center}
		\begin{minipage}{0.110\hsize}
			\centerline{\includegraphics[width=\hsize]{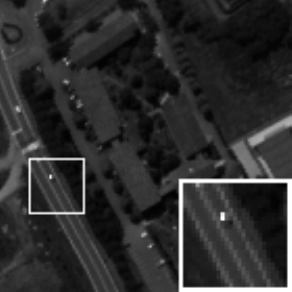}} 
		\end{minipage}	
		\begin{minipage}{0.110\hsize}
			\centerline{\includegraphics[width=\hsize]{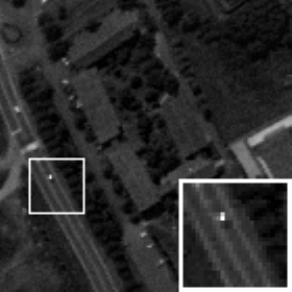}} 
		\end{minipage}
		\begin{minipage}{0.110\hsize}
			\centerline{\includegraphics[width=\hsize]{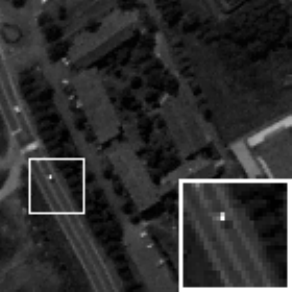}} 
		\end{minipage}
		\begin{minipage}{0.110\hsize}
			\centerline{\includegraphics[width=\hsize]{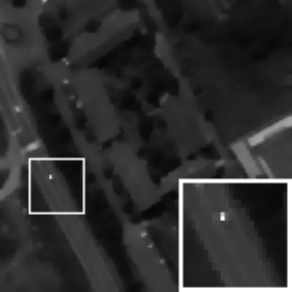}} 
		\end{minipage}
		\begin{minipage}{0.110\hsize}
			\centerline{\includegraphics[width=\hsize]{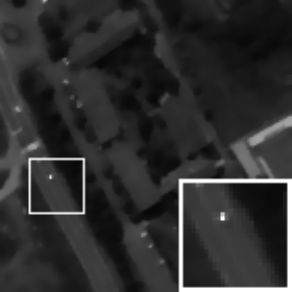}} 
		\end{minipage}
		\begin{minipage}{0.110\hsize}
			\centerline{\includegraphics[width=\hsize]{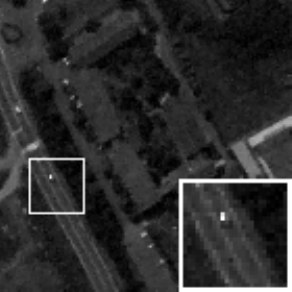}} 
		\end{minipage}
		\begin{minipage}{0.110\hsize}
			\centerline{\includegraphics[width=\hsize]{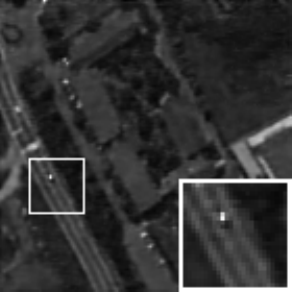}} 
		\end{minipage}
		\begin{minipage}{0.110\hsize}
			\centerline{\includegraphics[width=\hsize]{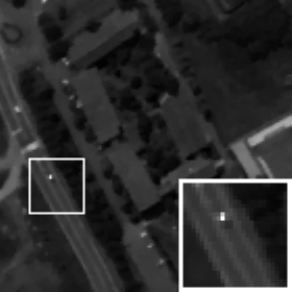}} 
		\end{minipage}
		\begin{minipage}{0.040\hsize}
			\centerline{\hspace{\hsize}} 
		\end{minipage}
		
		\vspace{1mm}
		
		\begin{minipage}{0.110\hsize}
			\centerline{\includegraphics[width=\hsize]{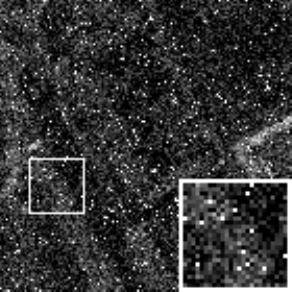}} 
		\end{minipage}
		\begin{minipage}{0.110\hsize}
			\centerline{\includegraphics[width=\hsize]{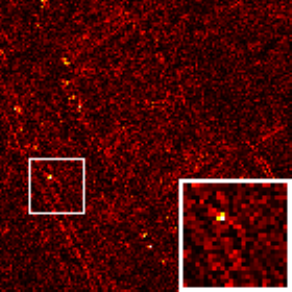}} 
		\end{minipage}
		\begin{minipage}{0.110\hsize}
			\centerline{\includegraphics[width=\hsize]{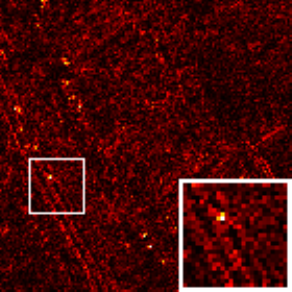}} 
		\end{minipage}
		\begin{minipage}{0.110\hsize}
			\centerline{\includegraphics[width=\hsize]{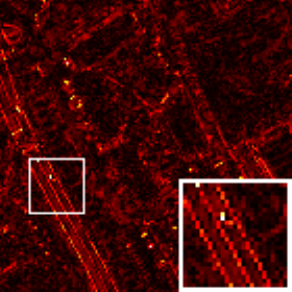}} 
		\end{minipage}
		\begin{minipage}{0.110\hsize}
			\centerline{\includegraphics[width=\hsize]{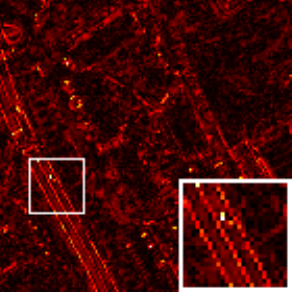}} 
		\end{minipage}
		\begin{minipage}{0.110\hsize}
			\centerline{\includegraphics[width=\hsize]{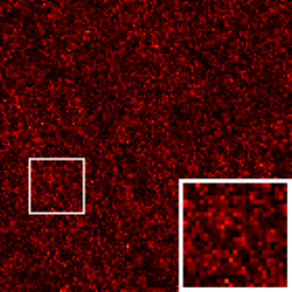}} 
		\end{minipage}
		\begin{minipage}{0.110\hsize}
			\centerline{\includegraphics[width=\hsize]{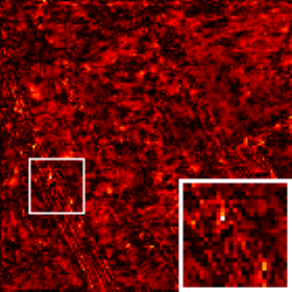}} 
		\end{minipage}
		\begin{minipage}{0.110\hsize}
			\centerline{\includegraphics[width=\hsize]{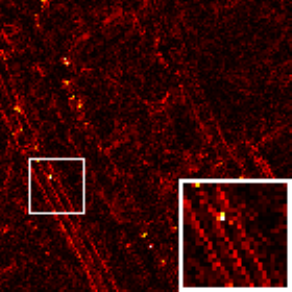}} 
		\end{minipage}
		\begin{minipage}{0.040\hsize}
			\centerline{\includegraphics[width=\hsize]{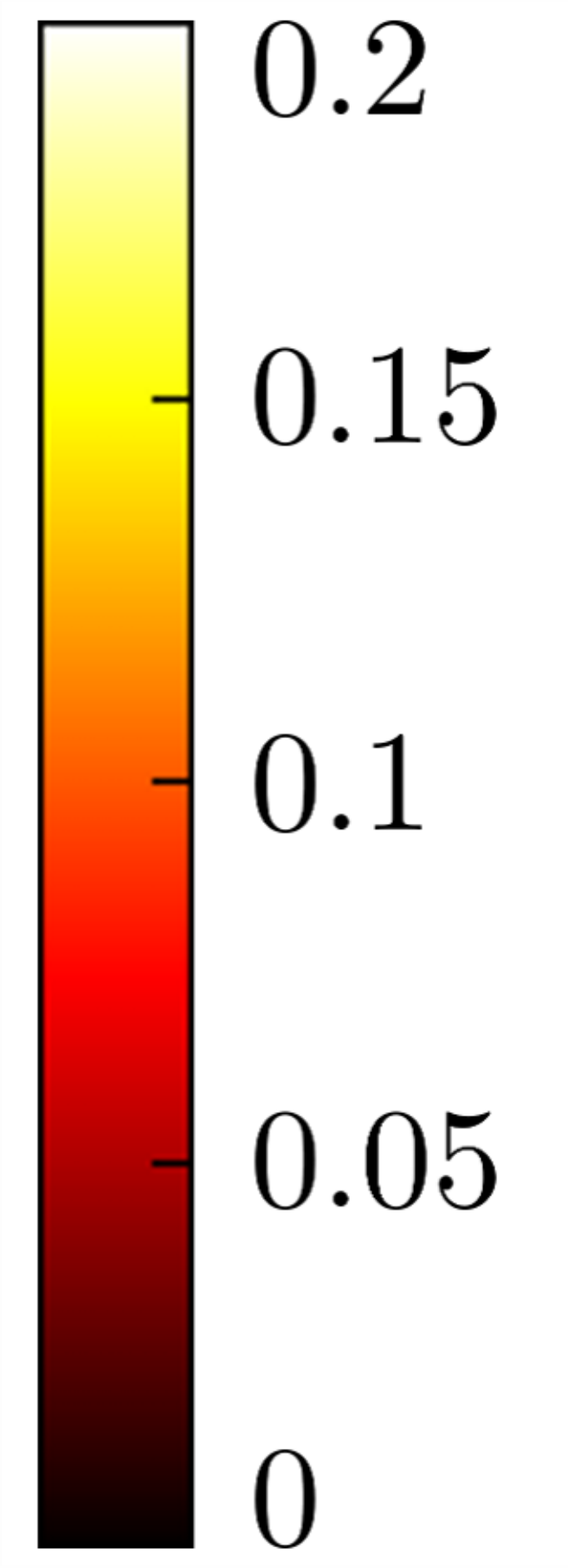}} 
		\end{minipage}
		
		\vspace{1mm}
		
		\begin{minipage}{0.110\hsize}
			\centerline{\small{Method}}
		\end{minipage}
		\begin{minipage}{0.110\hsize}
			\centerline{\small{SSTV~\cite{Aggarwal2016SSTV}}}
		\end{minipage}
		\begin{minipage}{0.110\hsize}
			\centerline{\small{$\llHTV$~\cite{Wang2021l0l1HTV}}}
		\end{minipage}
		\begin{minipage}{0.110\hsize}
			\centerline{\small{HSSTV1~\cite{Takeyama2020HSSTV}}}
		\end{minipage}
		\begin{minipage}{0.110\hsize}
			\centerline{\small{HSSTV2~\cite{Takeyama2020HSSTV}}}
		\end{minipage}
		\begin{minipage}{0.110\hsize}
			\centerline{\small{TPTV~\cite{Chen2023TPTV}}}
		\end{minipage}
		\begin{minipage}{0.110\hsize}
			\centerline{\small{QRNN3D~\cite{Wei2021QRNN3D}}}
		\end{minipage}
		\begin{minipage}{0.110\hsize}
			\centerline{\small{\textbf{GeoSSTV}}}
		\end{minipage}
		\begin{minipage}{0.040\hsize}
			\centerline{\hspace{\hsize}} 
		\end{minipage}

		\vspace{-0.5mm}

		\begin{minipage}{0.110\hsize}
			\centerline{\small{MPSNR}}
		\end{minipage}
		\begin{minipage}{0.110\hsize}
			\centerline{\small{33.04}}
		\end{minipage}
		\begin{minipage}{0.110\hsize}
			\centerline{\small{33.45}}
		\end{minipage}
		\begin{minipage}{0.110\hsize}
			\centerline{\small{33.62}}
		\end{minipage}
		\begin{minipage}{0.110\hsize}
			\centerline{\small{33.63}}
		\end{minipage}
		\begin{minipage}{0.110\hsize}
			\centerline{\small{\underline{33.68}}}
		\end{minipage}
		\begin{minipage}{0.110\hsize}
			\centerline{\small{32.57}}
		\end{minipage}
		\begin{minipage}{0.110\hsize}
			\centerline{\small{\textbf{34.24}}}
		\end{minipage}
		\begin{minipage}{0.040\hsize}
			\centerline{\hspace{\hsize}} 
		\end{minipage}

		\vspace{-0.5mm}

		\begin{minipage}{0.110\hsize}
			\centerline{\small{MSSIM}}
		\end{minipage}
		\begin{minipage}{0.110\hsize}
			\centerline{\small{0.8763}}
		\end{minipage}
		\begin{minipage}{0.110\hsize}
			\centerline{\small{0.8908}}
		\end{minipage}
		\begin{minipage}{0.110\hsize}
			\centerline{\small{0.8895}}
		\end{minipage}
		\begin{minipage}{0.110\hsize}
			\centerline{\small{0.8865}}
		\end{minipage}
		\begin{minipage}{0.110\hsize}
			\centerline{\small{0.8795}}
		\end{minipage}
		\begin{minipage}{0.110\hsize}
			\centerline{\small{\underline{0.8975}}}
		\end{minipage}
		\begin{minipage}{0.110\hsize}
			\centerline{\small{\textbf{0.9079}}}
		\end{minipage}
		\begin{minipage}{0.040\hsize}
			\centerline{\hspace{\hsize}} 
		\end{minipage}
		

	\end{center}
	
	\vspace{-3mm}
	\caption{Denoising results for Pavia University with the 20th band in Case 2, multiplied by 1.5 for visibility. The first column shows the ground-truth (upper) and the observed noisy image (lower). In the other columns, the upper row images are the restored results by each method, and the lower row images are the absolute differences between the ground-truth and each restored image.}
	\label{fig:result_image_PU_Case2}
\end{figure*}
\begin{figure*}[t]
	\begin{center}
		\begin{minipage}{0.110\hsize}
			\centerline{\includegraphics[width=\hsize]{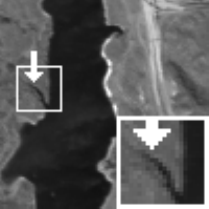}} 
		\end{minipage}	
		\begin{minipage}{0.110\hsize}
			\centerline{\includegraphics[width=\hsize]{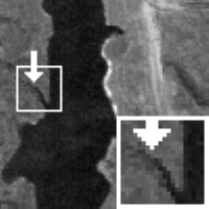}} 
		\end{minipage}
		\begin{minipage}{0.110\hsize}
			\centerline{\includegraphics[width=\hsize]{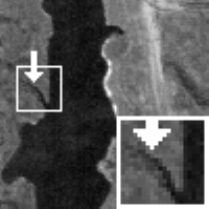}} 
		\end{minipage}
		\begin{minipage}{0.110\hsize}
			\centerline{\includegraphics[width=\hsize]{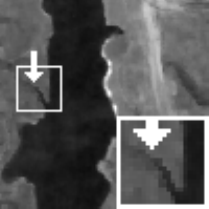}} 
		\end{minipage}
		\begin{minipage}{0.110\hsize}
			\centerline{\includegraphics[width=\hsize]{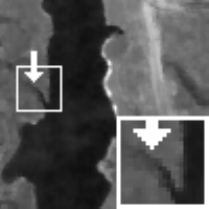}} 
		\end{minipage}
		\begin{minipage}{0.110\hsize}
			\centerline{\includegraphics[width=\hsize]{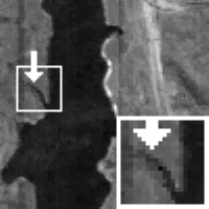}} 
		\end{minipage}
		\begin{minipage}{0.110\hsize}
			\centerline{\includegraphics[width=\hsize]{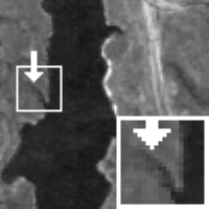}} 
		\end{minipage}
		\begin{minipage}{0.110\hsize}
			\centerline{\includegraphics[width=\hsize]{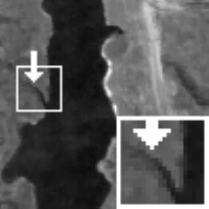}} 
		\end{minipage}
		\begin{minipage}{0.040\hsize}
			\centerline{\hspace{\hsize}} 
		\end{minipage}
		
		\vspace{1mm}
		
		\begin{minipage}{0.110\hsize}
			\centerline{\includegraphics[width=\hsize]{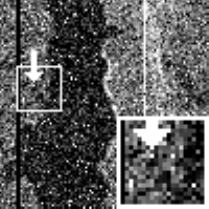}} 
		\end{minipage}
		\begin{minipage}{0.110\hsize}
			\centerline{\includegraphics[width=\hsize]{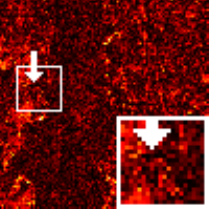}} 
		\end{minipage}
		\begin{minipage}{0.110\hsize}
			\centerline{\includegraphics[width=\hsize]{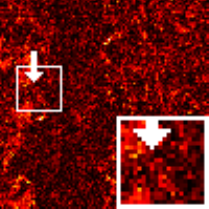}} 
		\end{minipage}
		\begin{minipage}{0.110\hsize}
			\centerline{\includegraphics[width=\hsize]{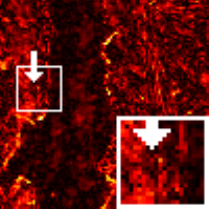}} 
		\end{minipage}
		\begin{minipage}{0.110\hsize}
			\centerline{\includegraphics[width=\hsize]{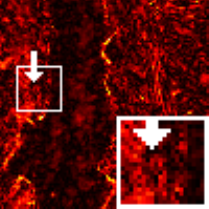}} 
		\end{minipage}
		\begin{minipage}{0.110\hsize}
			\centerline{\includegraphics[width=\hsize]{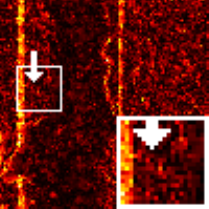}} 
		\end{minipage}
		\begin{minipage}{0.110\hsize}
			\centerline{\includegraphics[width=\hsize]{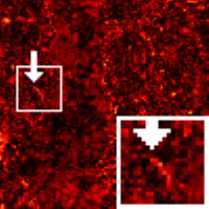}} 
		\end{minipage}
		\begin{minipage}{0.110\hsize}
			\centerline{\includegraphics[width=\hsize]{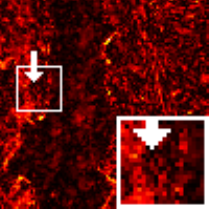}} 
		\end{minipage}
		\begin{minipage}{0.040\hsize}
			\centerline{\includegraphics[width=\hsize]{./result_image/colorbar_hot_r0.05-eps-converted-to.pdf}} 
		\end{minipage}
		
		\vspace{1mm}
		
		\begin{minipage}{0.110\hsize}
			\centerline{\small{Method}}
		\end{minipage}
		\begin{minipage}{0.110\hsize}
			\centerline{\small{SSTV~\cite{Aggarwal2016SSTV}}}
		\end{minipage}
		\begin{minipage}{0.110\hsize}
			\centerline{\small{$\llHTV$~\cite{Wang2021l0l1HTV}}}
		\end{minipage}
		\begin{minipage}{0.110\hsize}
			\centerline{\small{HSSTV1~\cite{Takeyama2020HSSTV}}}
		\end{minipage}
		\begin{minipage}{0.110\hsize}
			\centerline{\small{HSSTV2~\cite{Takeyama2020HSSTV}}}
		\end{minipage}
		\begin{minipage}{0.110\hsize}
			\centerline{\small{TPTV~\cite{Chen2023TPTV}}}
		\end{minipage}
		\begin{minipage}{0.110\hsize}
			\centerline{\small{QRNN3D~\cite{Wei2021QRNN3D}}}
		\end{minipage}
		\begin{minipage}{0.110\hsize}
			\centerline{\small{\textbf{GeoSSTV}}}
		\end{minipage}
		\begin{minipage}{0.040\hsize}
			\centerline{\hspace{\hsize}} 
		\end{minipage}

		\vspace{-0.5mm}

		\begin{minipage}{0.110\hsize}
			\centerline{\small{MPSNR}}
		\end{minipage}
		\begin{minipage}{0.110\hsize}
			\centerline{\small{\underline{34.83}}}
		\end{minipage}
		\begin{minipage}{0.110\hsize}
			\centerline{\small{34.55}}
		\end{minipage}
		\begin{minipage}{0.110\hsize}
			\centerline{\small{34.33}}
		\end{minipage}
		\begin{minipage}{0.110\hsize}
			\centerline{\small{34.33}}
		\end{minipage}
		\begin{minipage}{0.110\hsize}
			\centerline{\small{34.09}}
		\end{minipage}
		\begin{minipage}{0.110\hsize}
			\centerline{\small{32.07}}
		\end{minipage}
		\begin{minipage}{0.110\hsize}
			\centerline{\small{\textbf{35.18}}}
		\end{minipage}
		\begin{minipage}{0.040\hsize}
			\centerline{\hspace{\hsize}} 
		\end{minipage}

		\vspace{-0.5mm}

		\begin{minipage}{0.110\hsize}
			\centerline{\small{MSSIM}}
		\end{minipage}
		\begin{minipage}{0.110\hsize}
			\centerline{\small{0.9077}}
		\end{minipage}
		\begin{minipage}{0.110\hsize}
			\centerline{\small{0.8977}}
		\end{minipage}
		\begin{minipage}{0.110\hsize}
			\centerline{\small{\underline{0.9172}}}
		\end{minipage}
		\begin{minipage}{0.110\hsize}
			\centerline{\small{0.9169}}
		\end{minipage}
		\begin{minipage}{0.110\hsize}
			\centerline{\small{0.9076}}
		\end{minipage}
		\begin{minipage}{0.110\hsize}
			\centerline{\small{0.8660}}
		\end{minipage}
		\begin{minipage}{0.110\hsize}
			\centerline{\small{\textbf{0.9268}}}
		\end{minipage}
		\begin{minipage}{0.040\hsize}
			\centerline{\hspace{\hsize}} 
		\end{minipage}
		

	\end{center}
	
	\vspace{-3mm}
	\caption{Denoising results for Jasper Ridge with the 35th band in Case 5, multiplied by 2 for visibility. The first column shows the ground-truth (upper) and the observed noisy image (lower). In the other columns, the upper row images are the restored results by each method, and the lower row images are the absolute differences between the ground-truth and each restored image.}
	\label{fig:result_image_JR_Case5}
\end{figure*}
\begin{figure*}[t]
	\begin{center}
		\begin{minipage}{0.110\hsize}
			\centerline{\includegraphics[width=\hsize]{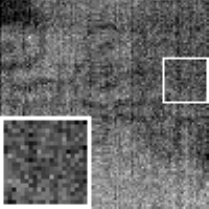}} 
		\end{minipage}	
		\begin{minipage}{0.110\hsize}
			\centerline{\includegraphics[width=\hsize]{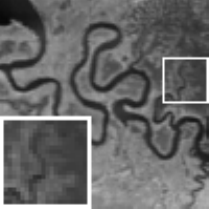}} 
		\end{minipage}
		\begin{minipage}{0.110\hsize}
			\centerline{\includegraphics[width=\hsize]{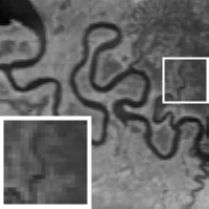}} 
		\end{minipage}
		\begin{minipage}{0.110\hsize}
			\centerline{\includegraphics[width=\hsize]{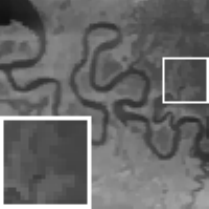}} 
		\end{minipage}
		\begin{minipage}{0.110\hsize}
			\centerline{\includegraphics[width=\hsize]{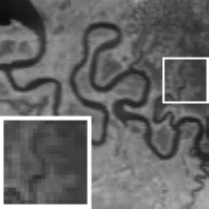}} 
		\end{minipage}
		\begin{minipage}{0.110\hsize}
			\centerline{\includegraphics[width=\hsize]{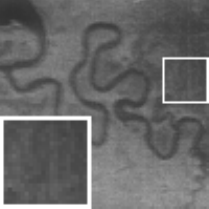}} 
		\end{minipage}
		\begin{minipage}{0.110\hsize}
			\centerline{\includegraphics[width=\hsize]{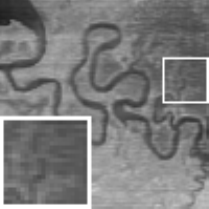}} 
		\end{minipage}
		\begin{minipage}{0.110\hsize}
			\centerline{\includegraphics[width=\hsize]{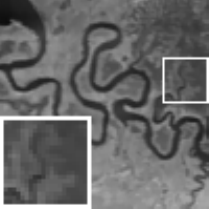}} 
		\end{minipage}

		\vspace{1mm}

		\begin{minipage}{0.110\hsize}
			\centerline{\includegraphics[width=\hsize]{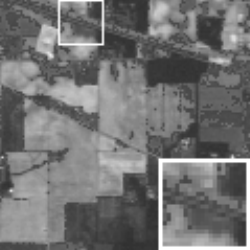}} 
		\end{minipage}	
		\begin{minipage}{0.110\hsize}
			\centerline{\includegraphics[width=\hsize]{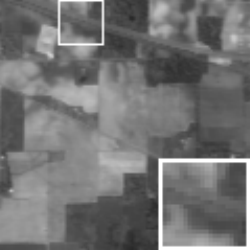}} 
		\end{minipage}
		\begin{minipage}{0.110\hsize}
			\centerline{\includegraphics[width=\hsize]{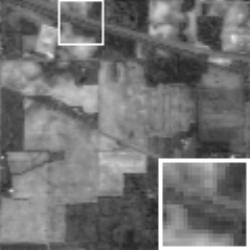}} 
		\end{minipage}
		\begin{minipage}{0.110\hsize}
			\centerline{\includegraphics[width=\hsize]{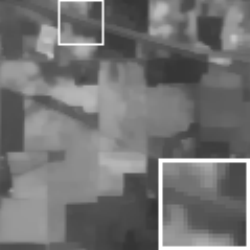}} 
		\end{minipage}
		\begin{minipage}{0.110\hsize}
			\centerline{\includegraphics[width=\hsize]{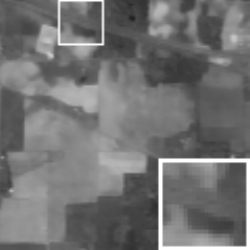}} 
		\end{minipage}
		\begin{minipage}{0.110\hsize}
			\centerline{\includegraphics[width=\hsize]{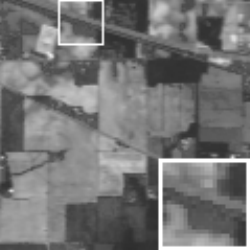}} 
		\end{minipage}
		\begin{minipage}{0.110\hsize}
			\centerline{\includegraphics[width=\hsize]{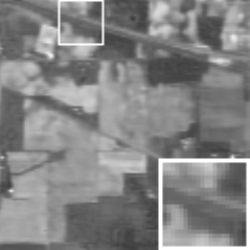}} 
		\end{minipage}
		\begin{minipage}{0.110\hsize}
			\centerline{\includegraphics[width=\hsize]{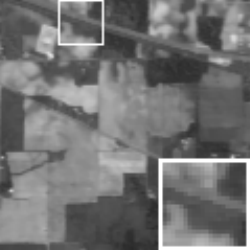}} 
		\end{minipage}
		
		\vspace{1mm}
		
		\begin{minipage}{0.110\hsize}
			\centerline{\small{Observation}}
		\end{minipage}
		\begin{minipage}{0.110\hsize}
			\centerline{\small{SSTV~\cite{Aggarwal2016SSTV}}}
		\end{minipage}
		\begin{minipage}{0.110\hsize}
			\centerline{\small{$\llHTV$~\cite{Wang2021l0l1HTV}}}
		\end{minipage}
		\begin{minipage}{0.110\hsize}
			\centerline{\small{HSSTV1~\cite{Takeyama2020HSSTV}}}
		\end{minipage}
		\begin{minipage}{0.110\hsize}
			\centerline{\small{HSSTV2~\cite{Takeyama2020HSSTV}}}
		\end{minipage}
		\begin{minipage}{0.110\hsize}
			\centerline{\small{TPTV~\cite{Chen2023TPTV}}}
		\end{minipage}
		\begin{minipage}{0.110\hsize}
			\centerline{\small{QRNN3D~\cite{Wei2021QRNN3D}}}
		\end{minipage}
		\begin{minipage}{0.110\hsize}
			\centerline{\small{\textbf{GeoSSTV}}}
		\end{minipage}


	\end{center}
	
    \vspace{-3mm}
    \caption{Denoising and destriping results on real noisy data for Suwannee with the 196th band (upper) and Indian Pines with the 32nd band(lower).}
    \label{fig:result_image_Real}
\end{figure*}

We adopt two HS image datasets which have different structures information. \\
\textit{1) Pavia University\footnotemark[2]:} This HS image was captured using a Reflective Optics System Imaging Spectrometer (ROSIS) sensor in Pavia, northern Italy. Pavia University consists of complex structures. The resolution of the original data is $610 \times 610$ pixels with 103 spectral bands per pixel. After removing several noisy bands and cropping the original data, we obtained the HS image with $120 \times 120$ pixels and 99 bands. \\
\textit{2) Jasper Ridge\footnote{\url{https://rslab.ut.ac.ir/data}}:} This HS image was captured using an Airborne Visible/Infrared Imaging Spectrometer (AVIRIS) sensor in a rural area of California, USA. Jasper Ridge consists of a large river in the center and fine structure on the left and right. The resolution of the original data is $512 \times 614$ pixels with 224 spectral bands per pixel. After removing several noisy bands and cropping the original data, we obtained the HS image with $100 \times 100$ pixels and 198 bands. \\


	
All the intensities of three HS images were normalized within the range $[0, 1]$.

HS images are often degraded by a mixture of various types of noise in real-world scenarios. Thus, in the experiments, we considered the following five cases of noise contamination:
\begin{itemize}
	\setlength{\leftskip}{18pt}
	\item[Case 1:] The observed HS image is contaminated by only white Gaussian noise with the standard deviation $\StanDevGauss = 0.1$.
	\item[Case 2:] The observed HS image is contaminated by Gaussian noise (Case~1) and additional salt-and-pepper noise with the rate $\RateSparse = 0.05$.
	\item[Case 3:] The observed HS image is contaminated by Gaussian noise (Case~1) and additional vertical stripe noise whose intensity is uniformly random in the range $[-0.5, 0.5]$ with the rate $\RateStripe = 0.05$.
	\item[Case 4:] The observed HS image is contaminated by Gaussian noise (Case~1) and additional deadline noise with the rate $\RateDeadline = 0.01$, where the stripe width chosen in the range $[1, 3]$. 
	\item[Case 5:] The observed HS image is contaminated by Gaussian noise (Case~1), salt-and-pepper noise (Case~2), stripe noise (Case~3), and deadline noise (Case~4).
\end{itemize}

The radii $\RadiusSparse$, $\RadiusStripe$, and $\RadiusFidel$ were set as follows:
\begin{align}
	\label{eq:RadiusSparse}
	\RadiusSparse &= \ParamsRadius \NumElement \bigl(0.5 \RateSparse + \MeanHSIObsv \CoverRateDeadline  \bigr), \\
	\label{eq:RadiusStripe}
	\RadiusStripe &= \ParamsRadius \tfrac{0.5 \NumElement \RateStripe (1 - \RateSparse) (1 - \CoverRateDeadline)}{2}, \\
	\label{eq:RadiusFidel}
	\RadiusFidel &= \ParamsRadius \sqrt{\StanDevGauss^2 \NumElement (1 - \RateSparse) (1 - \CoverRateDeadline)},
\end{align}
where $\MeanHSIObsv$ is the mean intensity of the observed HS image $\HSIObsv$, $\CoverRateDeadline$ is the coverage rate of deadline noise\footnote{
	Since overlapping deadlines make the effective coverage smaller than the deadline noise rate, the coverage rate of deadline noise is defined as
	\begin{equation*}
		\CoverRateDeadline = 1 - \exp(-\bar{w} \RateDeadline),
	\end{equation*}
	where $\bar{w}$ is the average width of the deadline noise.
}, and the parameter $\ParamsRadius$ was set depending on the number of constraints with nonzero radii: $0.98$ for Case~1 (one nonzero constraint), $0.95$ for Cases~2–4 (two nonzero constraints), and $0.90$ for Case~5 (three nonzero constraints).
The balancing parameter $\ParamBalance$ in \eqref{eq:GeoSSTV} was selected from $\{0.01, 0.03, 0.05\}$.
The stopping criterion of Alg.~1 were set as follows:
\begin{equation}
	\label{eq:StopCri_simulated}
	\frac{\| \HSIClean^{(\IndexAlg+1)} - \HSIClean^{(\IndexAlg)} \|_2}{\| \HSIClean^{(\IndexAlg)}\|_{2}} < 1.0 \times 10^{-5}.
\end{equation}

For the quantitative evaluation, we employed the mean peak signal-to-noise ratio (MPSNR):
\begin{equation}
	\label{eq:MPSNR}
	\mathrm{MPSNR} = \frac{1}{\NumBand} \sum_{\IndexBand=1}^{\NumBand} 10\log_{10}\frac{\NumVert \NumHori}{\|\HSIClean_{\IndexBand} - \bar{\HSIClean}_{\IndexBand}\|_{2}^{2}},
\end{equation}
and the mean structural similarity index (MSSIM)~\cite{Wang2004SSIM}:
\begin{equation}
	\label{eq:MSSIM}
	\mathrm{MSSIM} = \frac{1}{\NumBand} \sum_{\IndexBand=1}^{\NumBand} \mathrm{SSIM}(\HSIClean_{\IndexBand}, \bar{\HSIClean}_{\IndexBand}),
\end{equation}
where $\HSIClean_{\IndexBand}$ and $\bar{\HSIClean}_{\IndexBand}$ are the $\IndexBand$-th band of the ground true HS image $\HSIClean$ and the estimated HS image $\bar{\HSIClean}$, respectively.
Generally, higher MPSNR and MSSIM values are corresponding to better denoising performances.

\setcounter{subsubsection}{0}
\subsubsection{Quantitative Comparison}
\label{subsubsec:QuantitativeComparison}
Table~\ref{tab:MPSNR_MSSIM} shows MPSNRs and MSSIMs in the experiments on the HS image contaminated with simulated noise. The best and second best results are highlighted in bold and underlined, respectively.
QRNN3D shows higher performance for Pavia University than for Jasper Ridge. This can be attributed to the fine-tuning using the Pavia Centre data with similar spatial and spectral structures. The SSTV–LR hybrid method TPTV performs well in the cases without stripe noise, ranking second in terms of MPSNR in two cases. However, its effectiveness drops when stripe noise is present, as seen in Cases 3 and 5. In contrast, the existing TV-based methods, including SSTV, $\llHTV$, HSSTV1, and HSSTV2, exhibit relatively stable performance across different noise types. Among them, HSSTV2, which is an isotropic extension of TV, consistently shows superior MPSNRs and MSSIMs. On the other hand, GeoSSTV achieves the best MPSNRs and MSSIMs in all cases. Notably, the MSSIM values in Cases 4 and 5, where deadline noise is contaminated, are significantly higher than those of the existing TV-based methods. This demonstrates the superiority of GeoSSTV in restoring spatial structures even when pixel information is missing.

\subsubsection{Visual Quality Comparison for restored images}
Figs.~\ref{fig:result_image_PU_Case2} and \ref{fig:result_image_JR_Case5} show the results of HS image denoising and destriping. The lower row images are the absolute difference between the original image and each restored image.

Fig.~\ref{fig:result_image_PU_Case2} shows the denoising results for Pavia University in Case 2, i.e. under contamination by both Gaussian and sparse noise.
QRNN3D recovers most of the structures but suffers from noticeable blurring. SSTV, $\llHTV$, and TPTV leave residual noise along the road lines and throughout the image. In contrast, HSSTV1 and HSSTV2 effectively remove such noise owing to the direct promotion of spatial smoothness through the first-order difference term. However, in the enlarged view, the oblique road lines are lost. On the other hand, GeoSSTV removes noise while clearly retaining the road lines. This improvement can be attributed to its geometrically consistent formulation, which evaluates variations across all directions in the Euclidean manner, accurately preserving oblique edges.

Fig.~\ref{fig:result_image_JR_Case5} shows the denoising and destriping results for Jasper Ridge in Case 5, which is the most contaminated case with Gaussian, sparse, stripe, and deadline noise.
QRNN3D produces an overall reconstruction, but the groove along the river is truncated at the location indicated by the arrow. The restored image by TPTV exhibits residual stripe noise. This arises from the mischaracterization of stripe noise. In contrast, SSTV, $\llHTV$, HSSTV1, HSSTV2, and GeoSSTV succeed in removing stripe noise. This is thanks to the stripe noise characterization imposed by the second and third constraints in Eq.~\eqref{prob:GeoSSTV_denoising}. However, similar to the Pavia University results in Fig.~\ref{fig:result_image_PU_Case2}, SSTV and $\llHTV$ leave noise in the central river region. On the other hand, HSSTV1, HSSTV2, and our GeoSSTV successfully remove such noise, as can be seen in the difference images where the central river region appears closest to black.

\subsubsection{Parameter Analysis}
\label{subsubsec:ParamAnal}
\begin{figure}[t]
    \begin{center}
        \begin{minipage}{0.480\hsize}
            \centerline{\includegraphics[width=\hsize]{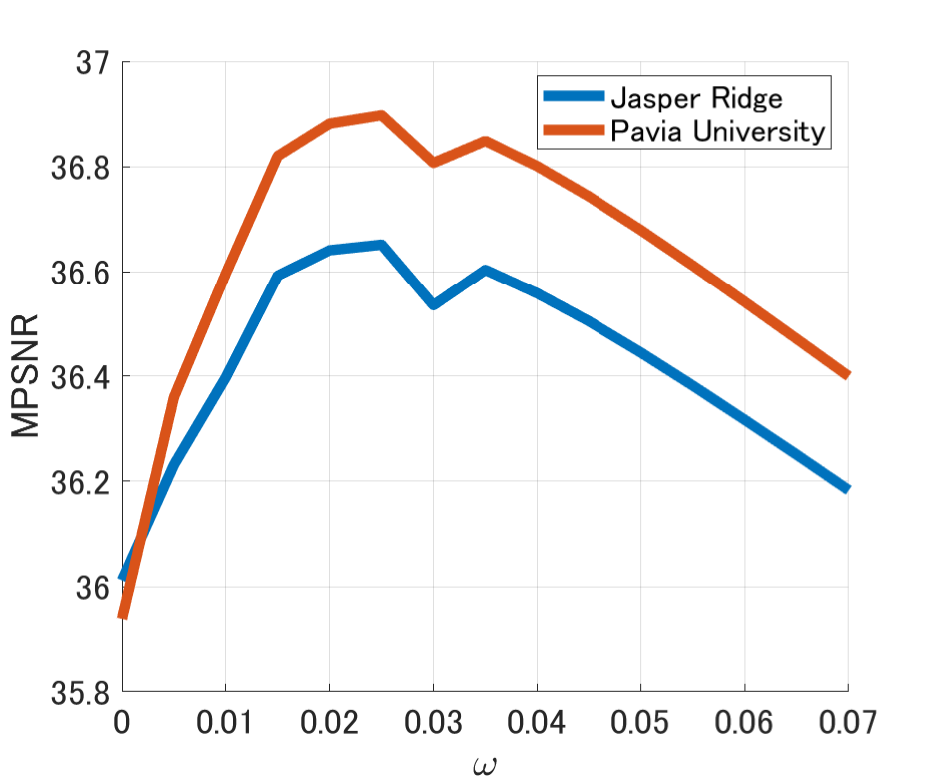}}
        \end{minipage}
        \begin{minipage}{0.480\hsize}
            \centerline{\includegraphics[width=\hsize]{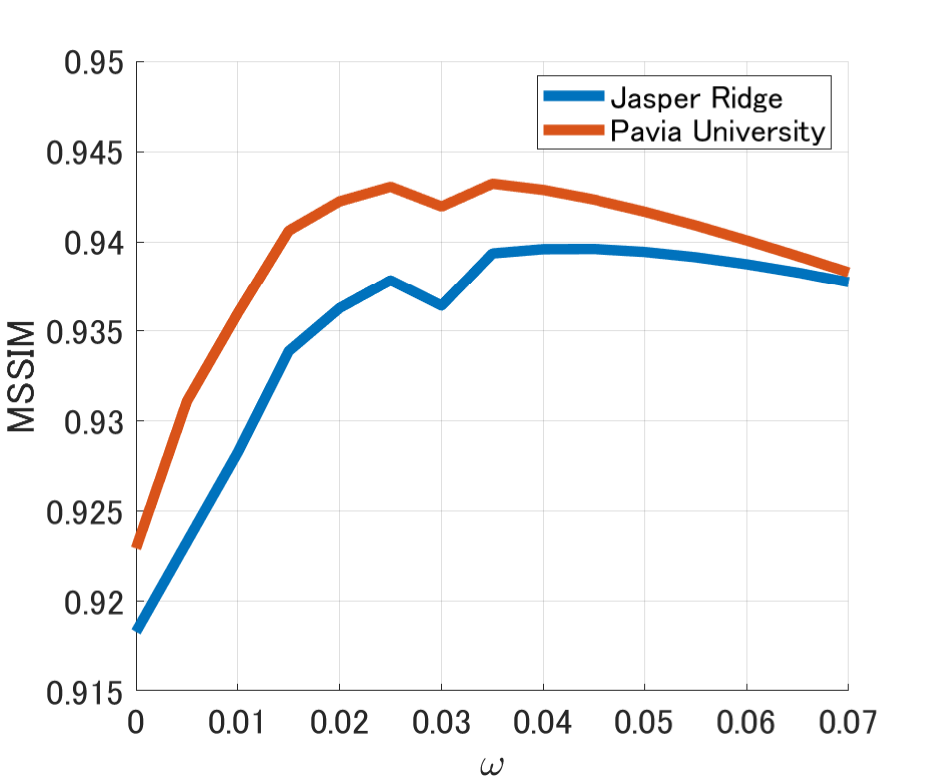}}
        \end{minipage}

        \begin{minipage}{0.480\hsize}
            \centerline{\small{(a)}}
		\end{minipage}
		\begin{minipage}{0.480\hsize}
			\centerline{\small{(b)}}
		\end{minipage}
    \end{center}
	
    \vspace{-3mm}
    \caption{Parameter analysis of the proposed method with respect to the balancing parameter $\ParamBalance$. (a) MPSNR versus $\ParamBalance$. (b) MSSIM versus $\ParamBalance$.}
    \label{fig:Param_Anal}
\end{figure}

The proposed method includes one key parameter: the balancing parameter $\ParamBalance$ between the first-order TV and the second-order TV terms. Fig.~\ref{fig:Param_Anal} shows the relationship between this parameter and the MPSNRs and MSSIMs.

When $\ParamBalance$ is set in the range of 0.01 to 0.05, the proposed method achieves high performance consistently for both MPSNR and MSSIM. Moreover, the variation across different images is slight. Based on this observation, we recommend setting $\ParamBalance$ within $[0.01, 0.05]$. In this paper, we selected $\ParamBalance$ from $\{0.01, 0.03, 0.05\}$ for all images and noise conditions.

\subsection{Real HS Image Experiment}
\label{subsec:RealHSIExpt}
We employed the following two datasets: \\
\textit{1) Indian Pines:} This HS image was captured using the AVIRIS sensor over the Indian Pines test site in North-western Indiana.
The resolution of the original data is $145 \times 145$ pixels, and each pixel has spectral information with 224 bands ranging from 400 nm to 2500 nm.
After removing several noisy bands and cropping the original data, we obtained the HS image with $120 \times 120$ pixels and 198 bands. \\
\textit{2) Suwannee:} This HS image was captured using a SpecTIR sensor over the Suwannee River Basin in Florida, USA.
The resolution of the original HS image is $1200 \times 320$ pixels, and each pixel has spectral information with 360 bands ranging from 400 nm to 2500 nm.
We cropped the HS image to $100 \times 100$ pixels and 360 bands. \\

All the intensities of both HS images were normalized within the range $[0, 1]$.
The balancing parameter $\ParamBalance$ was selected from $\{0.01, 0.03, 0.05\}$; as a result, the optimal value was $0.01$ for both images.
For the radii $\RadiusSparse$, $\RadiusStripe$, and $\RadiusFidel$, we adjusted them to appropriate values after empirically estimating the intensity of the noise in the real HS image.
Specifically, for the Indian Pines, $\RadiusSparse$, $\RadiusStripe$, and $\RadiusFidel$ were set to 200, 100, and 30, respectively, and for Suwannee, they were set to 800, 5000, and 100, respectively.
The stopping criterion of Alg.~1 were set as \eqref{eq:StopCri_simulated}.

Since no reference clean HS image is available, we compare the denoising performance using visual results.
Fig.~\ref{fig:result_image_Real} shows the HS image denoising and destriping results for Indian Pines and Suwannee.
For both real-noise datasets, TPTV leaves residual noise in the restored images. QRNN3D recovers spatial structures well, but the overall brightness is noticeably shifted toward higher intensity. HSSTV1 causes over-smoothing and loses fine spatial structures. In contrast, SSTV, $\llHTV$, HSSTV2, and GeoSSTV achieve sufficient noise removal while preserving detailed spatial structures. For Suwannee in particular, these methods successfully restore the narrow river structures severely corrupted by noise. Moreover, in the enlarged region of Indian Pines, GeoSSTV stands out by most clearly restoring the oblique lines, owing to its geometrically consistent formulation.



\section{Conclusion}
\label{sec:conclusion}

In this paper, we have proposed a new regularization method, named GeoSSTV, for HS image denoising and destriping. GeoSSTV integrates the first-order spatial TV and the second-order spatio-spectral TV within a geometrically consistent formulation that measures variations across all directions in a Euclidean manner. By leveraging this geometric consistency, GeoSSTV effectively removes noise and suppresses artifacts while preserving round structures and diagonal edges. We have formulated the denoising and destriping problem as a constrained convex optimization problem including GeoSSTV, and developed the optimization algorithm based on P-PDS. Experiments on HS images with simulated or real noise have demonstrated the superiority of GeoSSTV over existing methods.


\vfill\pagebreak



\end{document}